%% file: paper.tex
\documentclass[english,10pt,aps,prb,twocolumn,no,amsmath,amssymb,groupedaddress,nofootinbib]{revtex4-2}

\usepackage{silence}        
\usepackage{xparse}         

\usepackage[T1]{fontenc}    
\usepackage{lmodern}        
\usepackage[english]{babel} 
\usepackage[
    babel
    ]{microtype}            
\usepackage[all]{nowidow}   
\usepackage{xurl}           

\usepackage{mathtools}  
\usepackage{amstext}    
\usepackage{amssymb}    

\usepackage[
    separate-uncertainty=true,
    per-mode=symbol,
    range-phrase=\text{--},
    product-units=power
    ]{siunitx}          

\usepackage{csquotes}   

\usepackage{graphicx}   
\usepackage[
    inkscapearea=page
    ]{svg}              
\usepackage[
    off,                
    clean,              
    extractpath=svgdir
    ]{svg-extract}      
\usepackage{booktabs}   

\usepackage{mymath}
\usepackage{myphysics}

\usepackage[
    bookmarks=true,
    colorlinks,
    allcolors=blue
    ]{hyperref}                     
\usepackage[capitalize]{cleveref}   
\usepackage{orcidlink}
\usepackage{float} 

\allowdisplaybreaks[1]    

\graphicspath{{./diagrams/}{./diagrams/3p_ladders}{./plots/}}
\svgpath{{./diagrams/}{./diagrams/3p_ladders}}

\NewDocumentCommand{\includeinkscapefigure}{O{} m}{\includegraphics[#1]{#2}}

\NewDocumentCommand{\refcite}{O{}m}{Ref.~\cite[#1]{#2}}
\NewDocumentCommand{\refscite}{m}{Refs.~\cite{#1}}

\newcommand{\Title}{Three-particle ladders}
\newcommand{\Author}{Patrick Kappl}
\newcommand{\Keywords}{three-particle, 3-particle, chi2, AIM}

\hypersetup{
    pdfauthor = {\Author},
    pdftitle = {\Title},
    pdfkeywords = {\Keywords},
}

\begin{document}
    \title{Ladder equation for the three-particle vertex and its approximate solution}
    \author{Patrick Kappl\,\orcidlink{0000-0002-2475-9199}, Tin Ribic}
    \author{Anna Kauch\,\orcidlink{0000-0002-7669-0090}}
    \email{kauch@ifp.tuwien.ac.at}
    \author{Karsten Held\,\orcidlink{0000-0001-5984-8549}}

    \affiliation{Institute of Solid State Physics, TU Wien, 1040 Vienna, Austria}
    \date{\today}

    \input{abstract.tex}

    \maketitle
    \input{introduction.tex}
    \input{theoretical_background.tex}
    \input{three_particle_greens_function.tex}
    \input{three_particle_ladder.tex}

    % ...
    \input{conclusion_and_outlook.tex}
    \input{acknowledgments.tex}



    \bibliography{paper}
    \input{supplemental.tex}
\end{document}

%% file: abstract.tex
\begin{abstract}
    We generalize the three two-particle Bethe-Salpeter equations  to ten three-particle ladders. These equations are exact and yield the exact three-particle vertex, if we knew the three-particle vertex irreducible in one of the ten  channels. 
    However, as we do not have this three-particle irreducible vertex at hand, we approximate this building block for the ladder by the sum of two-particle irreducible vertices each connecting two fermionic lines. The comparison to the exact solution shows that this approximation is only good for rather weak interactions and even than only qualitatively --- at least for the  non-linear response function analyzed.
\end{abstract}

%% file: introduction.tex
\section{Introduction}
\label{sec:introduction}

Feynman diagrams are at the heart of the theory of strongly correlated electrons. Textbooks \cite{Abrikosov1975} study diagrams for one- and two-particle Green's functions which provide
 renormalizations (and life-times) of the quasiparticles and susceptibilities, respectively.  One-particle Green's functions and  self-energies also form the basis of
dynamical mean field theory (DMFT) \cite{Metzner1989,Georges1992a,Georges1996,DMFT25};
whereas two-particle Green's functions and vertices build the fundament of
diagramamtic extensions of DMFT \cite{Toschi2007,Rubtsov2008,RMPVertex}.

   Electronic correlations on the next level of three-particle  Feynman diagrams
are hitherto hardly explored. A calculation of selected three-particle contributions shows that these may become relevant  for diagrammatic extensions of DMFT \cite{Ribic2017a,Ribic2017b}\footnote{In Ref.~\cite{Hafermann2009}  only minor corrections have been reported, but this calculation was also in another parameter regime.}. 
Similarly, for the functional renormalization group
 the next level, hitherto
largely neglected, is  the three-particle level~\cite{Metzner2012}.
This calls for  a better understanding and for developing methods to actually calculate three-particle Green's functions and vertices more systematically.

Physics-wise, three-particle Green's functions are relevant for calculating Raman
and Hall responses~\cite{Jorio2011} as well as for
non-linear response functions  \cite{Kubo1957,Rostami2017,Rostami2021NLR,PhysRevB.103.195133,chi2-paper}. Traditionally 
such calculations have been restricted to calculating the three-particle correlator
from bubble contributions with quasiparticle-renormalizations of the Green's functions  but
without vertex corrections; or more recently by including a two-particle vertex between  the electron-hole pair of one of the bosonic fields/repsonses~\cite{Rostami2021NLR}. The three-particle Green's function has also recently beed used to obtain electronic spectra~\cite{Riva2022, Riva2023}   

In related research fields, 
three-particle physics is also relevant for  trions in semiconductors, which are typically treated as a three-particle Hamiltonian  \cite{Lampert1958,Combescot2004}. 
In quantum chromodynamics (QCD),
three-particle extensions of the Bethe-Salpeter equations, the so-called Faddeev equations \cite{Faddeev1961}, are employed for solving the three-particle problem of e.g.\ quarks forming protons or neutrons. Approximations such as the rainbow-ladder truncation \cite{Sanchis2013} have been developed. 
The Faddeev equations have also been employed
in nuclear physics to study the propagation of two holes and 
a particle interacting with each other through phonons~\cite{Barbieri2001}. A notable difference is that in solid-state physics our three particles or holes propagate in a background of many more electrons with which they interact {strongly}. This requires more
involved Feynman diagrams where we can have at a given time
{many} more (or less)  than three particle or hole excitations. A mapping onto a
three-particle Hamiltonian is no longer possible.

In this paper, we derive a three-particle analog of the Bethe-Salpeter equation. Irreducibility on the three-particle level is more involved and leads to additional one-particle reducible diagrams that need to be considered for the nine particle-particle-hole channels but not for the particle-particle-particle channel. Further, we solve this
generalized Bethe-Salpeter ladder for an approximate  irreducible three-particle vertex that consists only
of the two-particle vertices connecting all possible pairs of particle (hole) lines. To fulfill crossing symmetries of the vertex, contributions from multiple channels need to be considered. The thus approximated three-particle vertex is then compared to the exact one for an Anderson impurity model (AIM) at DMFT self-consistency. For this AIM, the vertex is local and can be calculated exactly.


The outline of the paper is as follows: Section \ref{sec:theoretical-background} sets the stage, defining the AIM and recalling some essentials for the two-particle diagrammatics. Section \ref{sec:3p-greens-function} defines the three-particle Green's function and vertex, as well as our frequency and spin notation. Section \ref{sec:3p-ladder} is the main part of this paper. Here, we derive  the three-particle Bethe-Salpeter-like ladder equations in Section~\ref{ssec:3p-bse}  and develop an approximation in terms of two particle irreducible vertices in Section~\ref{ssec:approximate-3p-ladder}. Section~\ref{sec:numerical-results} presents numerical results for the latter and compares them to exact ones.  Finally, Section \ref{sec:conclusion-and-outlook} summarizes our results and provides a brief outlook. 

%% file: theoretical_background.tex
\section{Theoretical background}
\label{sec:theoretical-background}
 In this Section, we define the AIM, the two-particle Green's function and vertex. We further recall the parquet and Bethe-Salpeter ladder equation as well as the crossing symmetry, which is very helpful for later understanding the extensions to three particles.
 
 \input{models_and_methods}

\subsection{Two-particle Green's function}
\label{ssec:g2}

\begin{figure*}
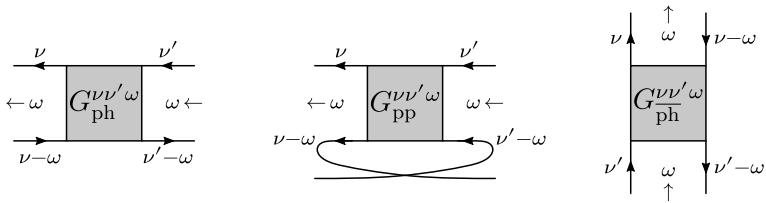

    \centering
    \includeinkscapefigure{G2FrequencyNotations}
    \caption{The same two-particle Green's function in the three frequency
             notations form \cref{tab:g2-frequency-notations}.}
    \label{fig:g2-frequency-notations}
\end{figure*}

The two-particle Green's function $\gf[2]$ is defined as
\begin{equation}
    \gf[2]_{1234}(\tau_1, \tau_2, \tau_3, \tau_4) \Def (-1)^2 \tev{
        \cq_1(\tau_1) \cqdag_2(\tau_2) \cq_3(\tau_3) \cqdag_4(\tau_4)}
    \label{eq:g2-tau}
\end{equation}
where $\mathrm T$ is the Wick {imaginary}-time ordering operator \cite{Abrikosov1975}, $\tau_1, \ldots, \tau_4$ are imaginary time variables, and where we use compound indices $1$ to $4$ for all quantum numbers such as spin, orbitals, and sites or momenta. The Fourier transformation from imaginary time  to Matsubara frequencies $\nu_i$  can be rewritten {by explicitly including} energy conservation as follows:
\begin{equation}
    \begin{split}
        \gf[2]_{1234}^{\nu_1 \nu_2 \nu_3 \nu_4}
            & := \iiiint_0^\invtemp \!\!  \dl{\tau_1, \tau_2, \tau_3, \tau_4} \,
            \gf[2]_{1234}(\tau_1, \tau_2, \tau_3, \tau_4) 
            \\
            & \phantom{\iiiint_0^\invtemp  \; \dl{\tau_1,}  }
            \times
            \ee^{\ii
 (\nu_1 \tau_1 - \nu_2 \tau_2 + \nu_3 \tau_3 - \nu_4 \tau_4)}
        \\
        & = \invtemp \delta^{\nu_1 - \nu_2 + \nu_3 - \nu_4, 0}
            \! \iiint_0^\invtemp  \!\!   \dl{\tau'_1, \tau'_2, \tau'_3} \,  \\ 
            & \phantom{=\beta} 
            \gf[2]_{1234}(\tau'_1, \tau'_2, \tau'_3)
            \ee^{\ii (\nu_1 \tau'_1 - \nu_2 \tau'_2 + \nu_3 \tau'_3)}
        \\
        & \fed \invtemp \delta^{\nu_1 - \nu_2 + \nu_3 - \nu_4, 0}
            \gf[2]_{1234}^{\nu_1 \nu_2 \nu_3}.
    \end{split}
    \label{eq:g2-nu-tau}
\end{equation}
With this we can infer the dimension (indicated in the following by $[\cdots]$) of the two-particle Green's function in imaginary
time and Matsubara space:
\begin{align}
    \dimension{\gf(\tau_1, \tau_2, \tau_3, \tau_4)}
        & = \dimension{\gf(\tau_1, \tau_2, \tau_3, \tau_4)} = 1,
    \\
    \dimension{\gf^{\nu_1 \nu_2 \nu_3 \nu_4}} & = \dimension{\tau}^4,
    \\
    \dimension{\gf^{\nu_1 \nu_2 \nu_3}} & = \dimension{\tau}^3.
\end{align}
For the two-particle Green's function, we can also define new combinations $\omega$, $\nu$, $\nu'$  of the three independent frequencies $\nu_1$, $\nu_2$, $\nu_3$ as defined in \cref{tab:g2-frequency-notations} and \cref{fig:g2-frequency-notations}.
The Bethe-Salpeter equation then becomes diagonal in $\omega$ for the particle-hole (ph), transversal-particle-hole ($\phb$),  and particle-particle (pp) channel,
see~\cref{eq:bse} below.

\begin{table}
    \centering
    \begin{tabular}{@{}lcccc@{}}
        \toprule
        Channel & $\nu_1$ & $\nu_2$ & $\nu_3$ & $\nu_4$
        \\
        \midrule
        $\ph$ & $\nu$ & $\nu - \omega$ & $\nu' - \omega$ & $\nu'$
        \\
        $\pp$ & $\nu$ & $\omega - \nu'$ & $\omega - \nu$ & $\nu'$
        \\
        $\phb$ & $\nu$ & $\nu'$ & $\nu' - \omega$ & $\nu - \omega$
        \\
        \bottomrule
    \end{tabular}
    \caption{Frequency notations for the $\ph$, $\pp$, and $\phb$
             channel.}
    \label{tab:g2-frequency-notations}
\end{table}

For the sake of completeness let us also define the one-particle Green's function and its Fourier transform:
\begin{equation}
    \gf[1]_{12}(\tau_1, \tau_2) \Def (-1)  \tev{
        \cq_1(\tau_1) \cqdag_2(\tau_2) }
    \label{eq:g1-}
\end{equation}

\begin{equation}
    \begin{split}
        \gf[1]_{12}^{\nu_1 \nu_2}
            & := \iint_0^\invtemp \!\!  \dl{\tau_1, \tau_2} \,
            \gf[1]_{12}(\tau_1, \tau_2)
            \ee^{\ii (\nu_1 \tau_1 - \nu_2 \tau_2)}\\
     & \fed \invtemp \delta^{\nu_1 - \nu_2 , 0}
            \gf[1]_{12}^{\nu_1}.
    \end{split}
    \label{eq:g1-nu-tau}
\end{equation}

\subsection{Full two-particle vertex}
\label{ssec:f2}

In the following we use skeleton diagrams
that are defined in terms of the full (interacting)
one-particle  Green's function and thus
must not contain one-particle
insertions (i.e., no internal parts that can be separated completely from the rest of the diagram by cutting two
Green's function lines)
\cite{Abrikosov1975}.
Furthermore, terms of the two-particle
Green's function that actually connect the two in- and outgoing
particle lines can be collected into an interaction vertex, which we call the full
two-particle vertex $\fv[2]$. In the following
we usually drop the particle-ness index (here "2") 
if it can be inferred from the (number of) arguments. With this the whole series expansion
can be decomposed into vertex contributions and disconnected Green's function lines (cf.~\cref{fig:g2-decomposition}):
\begin{figure*}
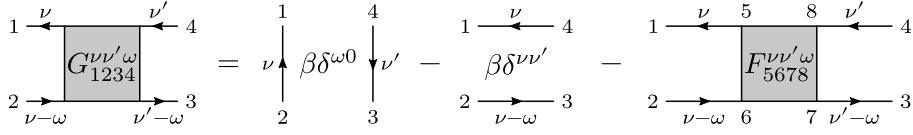

    \centering
    \includeinkscapefigure{G2Decomposition}
    \caption{Decomposition of the two-particle Green's function into two disconnected
             terms and a connected term, introducing the full two-particle vertex $\fv$}
    \label{fig:g2-decomposition}
\end{figure*}
\begin{equation}
    \begin{split}
        \gf_{1234}^{\nu \nu' \omega} = {}
        & \invtemp \delta^{\omega 0} \gf_{12}^{\nu} \gf_{34}^{\nu'}
            - \invtemp \delta^{\nu \nu'} \gf_{14}^{\nu} \gf_{32}^{\nu - \omega} \\ & 
            - \gf_{15}^{\nu} \gf_{62}^{\nu - \omega} \fv_{5678}^{\nu \nu' \omega}
            \gf_{37}^{\nu'} \gf_{84}^{\nu' - \omega},
    \end{split}
    \label{eq:g2-decomposition}
\end{equation}
Here, the sign of the vertex is just convention. Our choice leads
to the following weak coupling (first order) terms of $\fv$ for a local interaction:
\begin{equation}
    \begin{split}
    \lim_{\interaction \to 0^+} \fv_{\uparrow\uparrow\downarrow\downarrow} \equiv & \lim_{\interaction \to 0^+} \fv_{\ud}
        = +\interaction, \\
    \lim_{\interaction \to 0^+} \fv_{\uparrow\downarrow\downarrow\uparrow} \equiv &\lim_{\interaction \to 0^+} \fv_{\udb}
        = -\interaction.
            \end{split}
\end{equation}
From \cref{eq:g2-decomposition} we can also deduce the dimension of the full vertex to be
$    \dimension{\fv^{\nu \nu' \omega}} = \dimension{\tau}^{-1},
$
consistent with the bare interaction $\interaction$ which also has the
dimension of energy or inverse time.

\subsection{Irreducible two-particle vertices}
\label{ssec:irreducible-2p-vertices}
A further step is to classify the two-particle diagrams in terms of
their reducibility.
{For two-particle interactions, all fermionic two-particle diagrams are inherently one-particle irreducible
(1PI); i.e., it is not possible to separate a two-particle diagram by cutting one (single-particle) Green's function line (otherwise the conservation of the number of fermions would be violated)}. In contrast, three-particle diagrams discussed later will not be necessarily 1PI.

On the two-particle level, of course, we also need to think about two-particle
reducibility. This means that we check if diagrams can be separated into two disconnected
parts by cutting two Green's function lines.
{If not, we call the diagram} fully two-particle irreducible. {If yes, the diagram is two-particle reducible.} 
 We can further refine the classification of reducible diagrams by {looking at} pairs of external points which stay
connected: a
particle--hole pair that runs horizontally in \cref{fig:g2-frequency-notations}, a particle--hole pair that runs vertically, or
a particle--particle pair. These are the three channels ($\ph$, $\phb$, $\pp$) we already
mentioned in \cref{ssec:g2} when introducing the different frequency notations. 
While the frequency notations are just different parametrizations for the same object, convenient for evaluating Bethe-Salpeter equations (see~Eqs,~\eqref{eq:bse}) in the respective channels, the diagrams reducible in one channel are different from the diagrams reducible in another channel.  Moreover, it turns out every
two-particle diagram can only be reducible in  one of the three channels
\cite{Bickers1989,Rohringer2012,Bickers2004}. This allows us to uniquely split up the full
two-particle vertex $\fv$ into four terms
\begin{equation}
    \fv_{1234} = \fiv_{1234} + \rd_{\ph,1234} + \rd_{\phb,1234} + \rd_{\pp,1234},
    \label{eq:parquet-equation}
\end{equation}
where $\fiv$ is the fully irreducible two-particle vertex, and $\rd_{\ph}$, $\rd_{\phb}$, and $\rd_{\pp}$ are reducible vertices in $\ph$, $\phb$, and $\pp$ channels. \cref{eq:parquet-equation} is
called the parquet equation or parquet decomposition~\cite{Bickers2004}.

Let us now briefly talk about the crossing symmetry
(i.e., the symmetry under swapping the two in- and out-going legs) of the different terms in
\cref{eq:g2-decomposition,eq:parquet-equation}.
This symmetry will become relevant later for three-particle diagrams. Since the full two-particle Green's
function is crossing symmetric and the two disconnected terms in
\cref{eq:g2-decomposition} each map to the other one under crossing, the full
two-particle vertex is crossing symmetric as well. Things are similar for the components
of the parquet decomposition in \cref{eq:parquet-equation}. When swapping the two in- or
outgoing legs, the $\ph$ channel maps onto the $\phb$ channel and vice versa. The fully
irreducible vertex $\fiv$ never maps to any reducible diagram under exchange of external legs, and $\rd_{\pp}$ cannot map to $\fiv$ for the same reason.
Therefore, these two components are crossing symmetric on their own. Putting everything
together we have the following crossing symmetries:
\begin{equation}
    \begin{split}
        \fv_{1234} & = -\fv_{3214} = -\fv_{1432},
        \\
        \fiv_{1234} & = -\fiv_{3214} = -\fiv_{1432},
        \\
        \rd_{\ph,1234} & = -\rd_{\phb,3214} = -\rd_{\phb,1432},
        \\
        \rd_{\phb,1234} & = -\rd_{\ph,3214} = -\rd_{\ph,1432},
        \\
        \rd_{\pp,1234} & = -\rd_{\pp,3214} = -\rd_{\pp,1432}.
    \end{split}
    \label{eq:crossing-relations}
\end{equation}

\Cref{eq:parquet-equation} is not the only useful decomposition of the full two-particle
vertex $\fv$, regarding reducibility. Instead of working with the fully irreducible vertex
$\fiv$ we can look at one channel at a time and define vertices $\iv$ that are only
irreducible {in each of} them:
\begin{equation}
    \iv_r = \fv - \rd_r, \quad r \in \{\ph, \phb, \pp\}.
\end{equation}
With this we are back to a binary property for each channel $r$ (each diagram is part of either $\iv_r$ or $\rd_r$). From every reducible diagram in $\rd_r$ we can cut off the
left-most irreducible part and again be left with the whole series of diagrams for the
full vertex $\fv$. This leads to the so-called Bethe--Salpeter
equations \cite{bethe-salpeter}
\begin{equation}
    \begin{split}
        \fv_{1234} & = \iv_{\ph,1234}
            + \sum_{5678} \iv_{\ph,1256} \gf_{67} \gf_{85} \fv_{7834},
        \\
        \fv_{1234} & = \iv_{\phb,1234}
            - \sum_{5678} \iv_{\phb,1654} \gf_{67} \gf_{85} \fv_{7238},
        \\
        \fv_{1234} & = \iv_{\pp,1234}
            + \frac{1}{2} \sum_{5678} \iv_{\pp,1836} \gf_{67} \gf_{85} \fv_{7254},
    \end{split}
    \label{eq:bse}
\end{equation}
where the different signs between the $\ph$ and $\phb$ channels respect the crossing
relations in \cref{eq:crossing-relations} and the factor $1/2$ in the $\pp$ channel comes
from the indistinguishability of the particles \cite{Bickers2004} (for a detailed explanation of why the
factor of $1/2$ is necessary cf.\ the discussion after \cref{eq:connected-ladder} in
\cref{ssec:approximate-3p-ladder}).

%% file: models_and_methods.tex

\subsection{Anderson impurity model}
\label{sec:aim}

The AIM, introduced in 1961 by P.~W.~Anderson to describe
magnetic impurities in metals \cite{Anderson1961}, has become a popular
model in the field of correlated many-electron systems. Arguably, this is because even
though its Hamiltonian, which is described below, is rather simple it already shows
effects of strong correlation, more specifically the Kondo effect
\cite{Kondo1964,Schrieffer1966}. Another important aspect is that its solution is
connected to local correlation functions of more complex lattice Hamiltonians in  DMFT \cite{Georges1996} and its diagrammatic extensions \cite{RMPVertex}.  For an in-depth review of the physics of the AIM and the Kondo problem see
\refscite{Hewson1993,Coleman2015}.

The Hamiltonian of the AIM  reads
\begin{equation}
        \hamiltonian[AIM]  = 
            \energy \nq + \interaction \nq_\up \nq_\down
            + \sum_{k, \sigma} \energy^{\vphantom{\dagger}}_k
            \cqdag_{k \sigma} \cq_{k \sigma}
            \!+\! \sum_{k, \sigma} 
            \left(
                \hybridization^{\vphantom{\dagger}}_k
                \fqdag_{\sigma\vphantom{k}} \cq_{k \sigma}
               \!+
               \mathrm{h.c.}
                \right)
                .
    \label{eq:aim-hamiltonian}
\end{equation}
The first two terms describe the impurity with one-particle energy $\energy$. Electrons on
the impurity are created and annihilated with $\fqdag_\sigma$ and $\fq_\sigma$,
respectively. As usual, the density operators are given by $\nq = \nq_\up + \nq_\down$,
with $\nq_\sigma = \fqdag_\sigma \fq_\sigma$. If the impurity is doubly occupied, the local
Coulomb interaction $\interaction$ comes into play. The third term models a bath of
noninteracting electrons with dispersion relation $\energy_k$ {in terms of} creation and
annihilation operators $\cqdag_{k \sigma}$ and $\cq_{k \sigma}$. The last {two} terms 
{describe} {hopping (hybridization)} from the bath to the impurity ($\hybridization_k$) and vice versa
({Hermitian conjugate;} $\hybridization^*_k$).

Here, we employ the AIM with a real, $k$-independent hybridization
$\hybridization_k \equiv \hybridization \in \realNumbers$ and with a bath density of states (DOS) given by DMFT self-consistency for the Hubbard model~\cite{Georges1996} with the same $U$, hopping $t\equiv 1$ and inverse temperature $\beta=1/T=10$ (we set the Boltzmann constant $k_B$ to $1$). The motivation behind this choice is that we are eventually interested in diagrammatic extensions of DMFT which have the local two- or, on the next level, three-particle vertices of such an AIM  as a starting point, see Ref.~\onlinecite{RMPVertex}. However, here we do not pursue further calculations of non-local correlations of the Hubbard model. For the purpose of this paper, we only calculate one-, two-, and three-particle Green's functions and vertices of an AIM whose bath is determined by DMFT self-consistency.


%% file: three_particle_greens_function.tex
\section{Three-particle Green's function}
\label{sec:3p-greens-function}

\begin{figure}
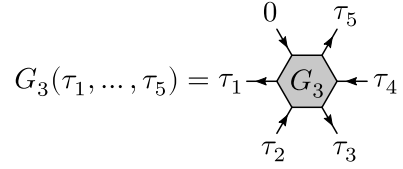

    \centering
    \includeinkscapefigure{G3}
    \caption{Diagrammatic representation of the three-particle Green's function in
             imaginary times}
    \label{fig:g3}
\end{figure}

{Let us now turn to the}
three-particle Green's function, {visualized in}  \cref{fig:g3} {and} defined as
\begin{multline}
    \gf[3]_{1 \dots 6}(\tau_1, \dots, \tau_5) \Def
    \\
    (-1)^3 \tev{\cq_1(\tau_1) \cqdag_2(\tau_2) \cq_3(\tau_3)
                \cqdag_4(\tau_4) \cq_5(\tau_5) \cqdag_6(0)},
    \label{eq:g3-tau}
\end{multline}
with imaginary times $\tau_i$ and all other parameters and quantum numbers condensed into compound indices $1 \dots 6$. Note that we directly used time translation invariance (energy conservation)  to
get rid of one time argument, analogous to  \cref{ssec:g2} for the two-particle Green's functions. 

The Fourier transformation to fermionic Matsubara frequencies $\nu_i = (2 n_i + 1) \pi /
\invtemp$, $n_i \in \integerNumbers$, is given by
\begin{eqnarray}
    \gf[3]_{1 \dots 6}^{\nu_1 \dots \nu_5} &=&
        \idotsint_0^\invtemp \dl{\tau_1} \dots \dl{\tau_5} \,
        \gf[3]_{1 \dots 6}(\tau_1, \dots, \tau_5)
       \\ &&  \times  \;
        \ee^{\ii (\nu_1 \tau_1 - \nu_2 \tau_2 + \nu_3 \tau_3 - \nu_4 \tau_4
                   + \nu_5 \tau_5)},\nonumber
    \label{eq:g3-nu-tau}
\end{eqnarray}
and the inverse transformation reads
\begin{eqnarray}
   \gf[3]_{1 \dots 6}(\tau_1, \dots, \tau_5) &=&
        \frac{1}{\invtemp^5} \sum_{\nu_1 \dots \nu_5}
        \gf[3]_{1 \dots 6}^{\nu_1 \dots \nu_5} 
       \\ && \nonumber \times 
       \ee^{-\ii (\nu_1 \tau_1 - \nu_2 \tau_2 + \nu_3 \tau_3 - \nu_4 \tau_4
                  + \nu_5 \tau_5)}.
    \label{eq:g3-tau-nu}
\end{eqnarray}
With that we can
infer the dimension of the three-particle Green's function in imaginary time and Matsubara
space:
\begin{align}
    \dimension{\gf[3]_{1 \dots 6}(\tau_1, \dots, \tau_5)} & = 1,
    \\
    \dimension{\gf[3]_{1 \dots 6}^{\nu_1 \dots \nu_5}} & = \dimension{\tau}^5.
\end{align}

From now on we drop the index $3$ if the three-particle-ness can be inferred from the
(number of) arguments.

\subsection{Frequency notations}
\label{ssec:frequency-notations}

In \cref{ssec:g2} we saw that there are three channels for the two-particle Green's
function, each with its own frequency notation. They are chosen such that the in- and
outgoing particle--particle or particle--hole pairs have a total energy of $\omega$.


As shown in Ref.~\onlinecite{chi2-paper}, generalizing this yields 15 different two-particle
frequency notations for three-particle diagrams. They are listed in
\cref{tab:g3-2p-frequency-notations}.

\begin{table}
    \centering
    \begin{tabular}{@{}lcccccc@{}}
        \toprule
        Channel & $\nu_{1}$ & $\nu_{2}$ & $\nu_{3}$ & $\nu_{4}$ & $\nu_{5}$ & $\nu_{6}$
        \\
        \midrule
        $\ph$ & $\nu_a$ & $\nu_a - \omega_a$ & $\nu_b$ & $\nu_b - \omega_b$ & $\nu_c$
            & $\nu_c - \omega_c$
        \\
        $\ph'$ & $\nu_a$ & $\nu_b - \omega'_b$ & $\nu_b$ & $\nu_c - \omega'_c$ & $\nu_c$
            & $\nu_a - \omega'_a$
        \\
        $\phb$ & $\nu_a$ & $\nu_c - \omegabar_c$ & $\nu_b$ & $\nu_a - \omegabar_a$
            & $\nu_c$ & $\nu_b - \omegabar_b$
        \\
        $\phab$ & $\nu_a$ & $\nu_b - \omega'_b$ & $\nu_b$
            & $\nu_a - \omegabar_a$ & $\nu_c$ & $\nu_c - \omega_c$
        \\
        $\phbb$ & $\nu_a$ & $\nu_a - \omega_a$ & $\nu_b$ & $\nu_c - \omega'_c$
        & $\nu_c$ & $\nu_b - \omegabar_b$
        \\
        $\phcb$ & $\nu_a$ & $\nu_c - \omegabar_c$ & $\nu_b$
            & $\nu_b - \omega_b $ & $\nu_c$ & $\nu_a - \omega'_a$
        \\
        \midrule
        $\pp_{24-13}$ & $\nu_a$ & $\nu_b$ & $\omega_a - \nu_a$ & $\omega_b - \nu_b$
            & $\nu_c$ & $\nu_c - \omega_c$
        \\
        $\pp_{26-13}$ & $\omega_b - \nu_b$ & $\nu_a$ & $\nu_b$ & $\nu_c - \omega'_c$
            & $\nu_c$ & $\omega_a - \nu_a$
        \\
        $\pp_{26-15}$ & $\omega_c - \nu_c$ & $\omega_a - \nu_a$ & $\nu_b$
            & $\nu_b - \omega_b$ & $\nu_c$ & $\nu_a$
        \\
        $\pp_{46-15}$ & $\nu_a$ & $\nu_b - \omega'_b$ & $\nu_b$ & $\omega_c - \nu_c$
            & $\omega_a - \nu_a$ & $\nu_c$
        \\
        $\pp_{46-35}$ & $\nu_a$ & $\nu_a - \omega_a$ & $\nu_b$ & $\nu_c$
            & $\omega_b - \nu_b$ & $\omega_c - \nu_c$
        \\
        $\pp_{24-35}$ & $\nu_a$ & $\omega_b - \nu_b$ & $\omega_c - \nu_c$ & $\nu_b$
            & $\nu_c$ & $\nu_a - \omega'_a$
        \\
        \midrule
        $\pp_{26-35}$ & $\nu_a$ & $\nu_b$ & $\omega_c - \nu_c$
            & $\nu_a - \omegabar_a$ & $\nu_c$ & $\omega_b - \nu_b$
        \\
        $\pp_{46-13}$ & $\omega_b - \nu_b$ & $\nu_c - \omegabar_c$ & $\nu_b$
            & $\omega_a - \nu_a$ & $\nu_c$ & $\nu_a$
        \\
        $\pp_{24-15}$ & $\nu_a$ & $\omega_c - \nu_c$ & $\nu_b$ & $\nu_c$
            & $\omega_a - \nu_a$ & $\nu_b - \omegabar_b$
        \\
        \bottomrule
    \end{tabular}
    \caption{The 15 different two-particle frequency notations of three-particle diagrams \cite{chi2-paper}.}
    \label{tab:g3-2p-frequency-notations}
\end{table}

\begin{figure}
    \centering
    \includeinkscapefigure[width=8cm]{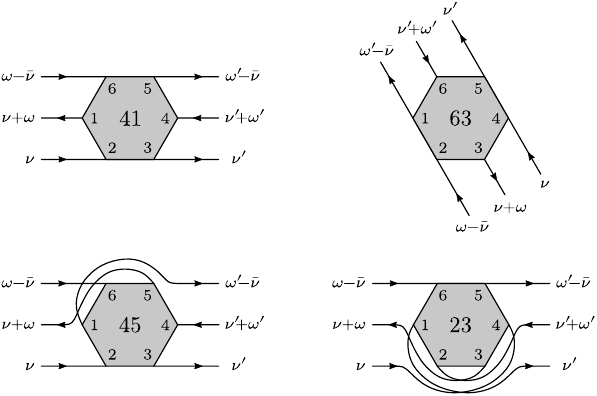}
    \caption{Diagrammatic representation of four of the ten three-particle frequency
             notations listed in \cref{tab:3p-notations}.}
    \label{fig:3p-notations}
\end{figure}

We can generalize this to three-particle frequency notations where we combine
either two particle lines and one hole line (pph) or three particle lines (ppp) such that
they have a total energy of $\bar{\nu}$. In the pph case there are $\binom{3}{2} \cdot
\binom{3}{1} = 9$ such combinations while in the ppp case there is only one for a total of
ten three-particle frequency notations. All of them are listed in \cref{tab:3p-notations}
while a diagrammatic representation is shown for four of them in \cref{fig:3p-notations}.
\begin{table}
    \begin{tabular}{@{}llcccccc@{}}
        \toprule
        Channel & & $\nu_1$ & $\nu_2$ & $\nu_3$ & $\nu_4$ & $\nu_5$ & $\nu_6$
        \\
        \midrule
        $(345)(612)$ & $41$
            & $\nu\! + \!\omega$     & $\nu$              & $\nu'$             
            & $\nu'\! + \!\omega'$   & $\omega'\! - \! \nubar$ & $\omega\! - \! \nubar$  
        \\
        $(123)(456)$ & $25$
            & $\nu'$             & $\nu'\! + \!\omega'$   & $\omega'\! - \! \nubar$ 
            & $\omega\! - \! \nubar$  & $\nu\! + \!\omega$     & $\nu$              
        \\
        $(561)(234)$ & $63$
            & $\omega'\! - \! \nubar$ & $\omega\! - \! \nubar$  & $\nu\! + \!\omega$     
            & $\nu$              & $\nu'$             & $\nu'\! + \!\omega'$   
        \\
        $(341)(652)$ & $45$
            & $\omega'\! - \! \nubar$ & $\nu$              & $\nu'$             
            & $\nu'\! + \!\omega'$   & $\nu\! + \!\omega$     & $\omega\! - \! \nubar$  
        \\
        $(361)(452)$ & $65$
            & $\omega'\! - \! \nubar$ & $\nu$              & $\nu'$             
            & $\omega\! - \! \nubar$  & $\nu\! + \!\omega$     & $\nu'\! + \!\omega'$   
        \\
        $(365)(412)$ & $61$
            & $\nu\! + \!\omega$     & $\nu$              & $\nu'$             
            & $\omega\! - \! \nubar$  & $\omega'\! - \! \nubar$ & $\nu'\! + \!\omega'$   
        \\
        $(325)(614)$ & $21$
            & $\nu\! + \!\omega$     & $\nu'\! + \!\omega'$   & $\nu'$             
            & $\nu$              & $\omega'\! - \! \nubar$ & $\omega\! - \! \nubar$  
        \\
        $(125)(634)$ & $23$
            & $\nu'$             & $\nu'\! + \!\omega'$   & $\nu\! + \!\omega$     
            & $\nu$              & $\omega'\! - \! \nubar$ & $\omega\! - \! \nubar$  
        \\
        $(145)(632)$ & $43$
            & $\nu'$             & $\nu$              & $\nu\! + \!\omega$     
            & $\nu'\! + \!\omega'$   & $\omega'\! - \! \nubar$ & $\omega\! - \! \nubar$  
        \\
        \midrule
        $(315)(642)$ & $\ppp$
            & $\omega'\! - \! \nu'$    & $\nu$               & $\nu'$              
            & $\omega\! - \! \nu$      & $-\omega'\! - \! \nubar$ & $-\omega\! - \! \nubar$  
        \\
        \bottomrule
    \end{tabular}
    \caption{The ten different three-particle frequency notations of three-particle
             diagrams. In our notation, the channel names in the first column (the two bracketed triplets) denote which particles and
             holes remain together. These triplets sum to a fermionic transfer frequency
             $\nubar$. The second column is a shorthand notation for these channel names; they are simply the middle numbers of the two triplets in the first
             column. For the $\pph$ channels (the first nine rows) they specify the
             particle running in the opposite direction. Diagrammatic representations for
             four of them are shown in \cref{fig:3p-notations}.}
    \label{tab:3p-notations}
\end{table}

With regard to the possible spin components there are only  three  independent spin
combinations for the six compound indices $1\cdots 6$ of Eq.~\ref{eq:g3-tau}: $\sigma_1 \dots \sigma_6 \in \{\up\up\up\up\up\up, \up\up\up\up\down\down,
\up\up\up\down\down\up\}$ which from now on we write more compactly as  $\uuu$, $\uud$,
$\ububd*$, respectively \cite{chi2-paper}.

Similar as for the two-particle Green's function in
Section~\ref{sec:theoretical-background}, one can define a full three-particle vertex
$\fv[3]$ which contains all diagrams that connect all incoming and outgoing lines  of
$\gf[3]$ in Fig.~\ref{fig:g3}\cite{chi2-paper} (additionally $\gf[3]$ contains disconnetced diagrams).

\begin{figure}
    \centering
    \includeinkscapefigure[width=\columnwidth]{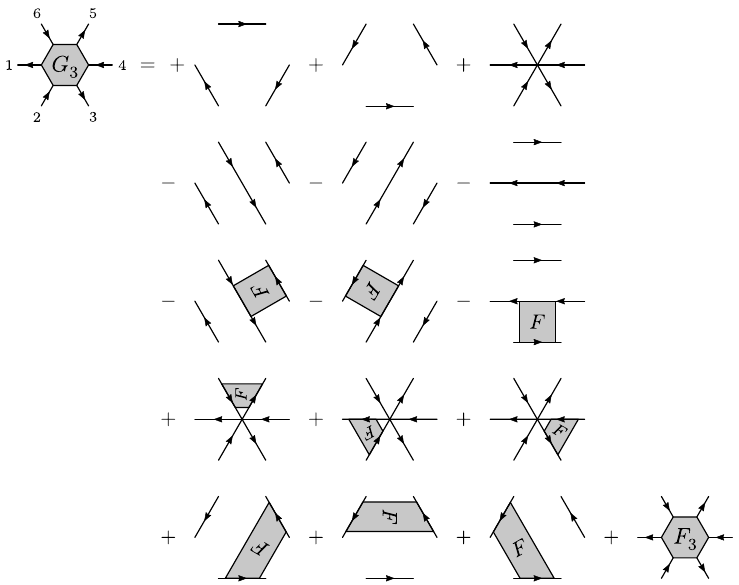}
    \caption{Diagrammatic representation of Eq. \eqref{eq:g3-decomposition}, the decomposition
             of the three-particle Green's function.}
    \label{fig:g3-decomposition}
\end{figure}

The decomposition of the three-particle Green's function is, cf.~\cref{fig:g3-decomposition}:
\begin{eqnarray}
      \!\!\!\!\! \lefteqn{\gf_{123456}=
        - \tev{\cqdag_1 \cq_2 \cqdag_3 \cq_4 \cqdag_5 \cq_6}} &&
        \\
       \;\; && \nonumber
        =  - \ctev{\cq_1 \cqdag_2} \ctev{\cq_3 \cqdag_4} \ctev{\cq_5 \cqdag_6}
        - \ctev{\cq_1 \cqdag_6} \ctev{\cq_3 \cqdag_2} \ctev{\cq_5 \cqdag_4}
        \\   
         \;\;  \;\; &&  \;\; \;\; \nonumber
        - \ctev{\cq_1 \cqdag_4} \ctev{\cq_3 \cqdag_6} \ctev{\cq_5 \cqdag_2}
        + \ctev{\cq_1 \cqdag_2} \ctev{\cq_3 \cqdag_6} \ctev{\cq_5 \cqdag_4}
        \\        
        \;\;  \;\; &&  \;\; \;\; \nonumber
        + \ctev{\cq_1 \cqdag_6} \ctev{\cq_3 \cqdag_4} \ctev{\cq_5 \cqdag_2}
        + \ctev{\cq_1 \cqdag_4} \ctev{\cq_3 \cqdag_2} \ctev{\cq_5 \cqdag_6}
        \\
        \;\;  \;\;  &&  \;\; \;\; \nonumber
        - \ctev{\cq_1 \cqdag_2} \ctev{\cq_3 \cqdag_4 \cq_5 \cqdag_6}
        - \ctev{\cq_3 \cqdag_4} \ctev{\cq_5 \cqdag_6 \cq_1 \cqdag_2}
        \\
      \;\;  \;\;    &&  \;\; \;\; \nonumber
        - \ctev{\cq_5 \cqdag_6} \ctev{\cq_1 \cqdag_2 \cq_3 \cqdag_4}
        + \ctev{\cq_1 \cqdag_4} \ctev{\cq_3 \cqdag_2 \cq_5 \cqdag_6} 
        \\
      \;\;  \;\;    &&  \;\; \;\; \nonumber
        + \ctev{\cq_3 \cqdag_6} \ctev{\cq_5 \cqdag_4 \cq_1 \cqdag_2}
        + \ctev{\cq_5 \cqdag_2} \ctev{\cq_1 \cqdag_6 \cq_3 \cqdag_4}
        \\
       \;\;  \;\;   &&  \;\; \;\; \nonumber
        + \ctev{\cq_1 \cqdag_6} \ctev{\cq_3 \cqdag_4 \cq_5 \cqdag_2}
        + \ctev{\cq_3 \cqdag_2} \ctev{\cq_5 \cqdag_6 \cq_1 \cqdag_4}
        \\
        \;\;  \;\;  &&  \;\; \;\;
        + \ctev{\cq_5 \cqdag_4} \ctev{\cq_1 \cqdag_2 \cq_3 \cqdag_6}
        -\ctev{\cq_1 \cqdag_2 \cq_3 \cqdag_4 \cq_5 \cqdag_6}
        .
    \label{eq:g3-decomposition}
\end{eqnarray}
where underlined expectation values denote fully connected diagrams. On the one-particle
level this means
\begin{equation}
    \gf_{12} = -\tev{\cq_1 \cqdag_2} = -\ctev{\cq_1 \cqdag_2},
    \label{eq:g1-ctev}
\end{equation}
and on the two-particle level we get
\begin{eqnarray}
    \gf_{1234} = \tev{\cq_1 \cqdag_2 \cq_3 \cqdag_4} &= &
        \ctev{\cq_1 \cqdag_2 \cq_3 \cqdag_4} \nonumber
        + \ctev{\cq_1 \cqdag_2} \ctev{\cq_3 \cqdag_4} \\ && 
        + (-1)^3 \ctev{\cq_1 \cqdag_4} \ctev{\cq_3 \cqdag_2},
    \label{eq:g2-ctev}
\end{eqnarray}
where comparison with \cref{eq:g2-decomposition} reveals that the four-point {fully connected} expectation
value is given by
\begin{equation}
    \ctev{\cq_1 \cqdag_2 \cq_3 \cqdag_4} =
        - \gf_{15} \gf_{62} \fv_{5678} \gf_{37} \gf_{84}.
    \label{eq:f2-ctev}
\end{equation}
Finally, the last term in \cref{eq:g3-decomposition} introduces the full three-particle
vertex $\fv[3]$. Its sign is chosen as
\begin{multline}
    -\ctev{\cq_1 \cqdag_2 \cq_3 \cqdag_4 \cq_5 \cqdag_6} = \\
        \gf_{1'1} \gf_{22'} \gf_{3'3} \fv[3]_{1'2'3'4'5'6'} \gf_{44'}\gf_{5'5}\gf_{66'}.
    \label{eq:f3-ctev}
\end{multline}

\subsection{Nonlinear response functions}
\label{ssec:nonlinear-response-functions}

\begin{figure}
    \centering
    \includeinkscapefigure[width=\columnwidth]{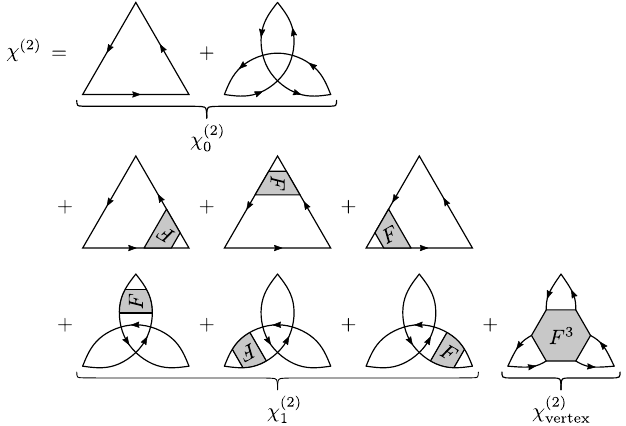}
    \caption{Diagrammatic representation of the decomposition of the second-order response
             function $\rf[2]$ into bare or bubble terms $\rf_0$, first-order terms
             $\rf_1$ and full vertex terms $\vrf$. Time and frequency labels are omitted
             to avoid clutter.}
    \label{fig:chi2-decomposition}
\end{figure}

As shown in Ref.~\onlinecite{chi2-paper} nonlinear response functions are the connected part ($\connected$)  of
three-particle correlators. Specifically, the second-order spin response functions
$\rf[2]_{\sigma_1 \sigma_2 \sigma_3}$ are defined as
\begin{equation}
    \rf[2]_{\sigma_1 \sigma_2 \sigma_3}(\tau_1, \tau_2) =
        \ctev{\nq_{\sigma_1}(\tau_1) \nq_{\sigma_2}(\tau_2) \nq_{\sigma_3}(0)},
\end{equation}
with the usual definition of the density operators $\nq_{\sigma} = \cqdag_{\sigma}
\cq_{\sigma}$; and the underline again denotes fully connected diagrams. Comparing this to \cref{eq:g3-tau} reveals that we can also write it in
terms of the connected part (conn) of the three-particle Green's function:
\begin{equation}
    \rf[2]_{\sigma_1 \sigma_2 \sigma_3}(\tau_1, \tau_2) =
        -\connected \gf_{\sigma_1 \sigma_3 \sigma_2}(
            \tau_1^{\vphantom +}, \tau_1^+, 0, 0^+, \tau_2^{\vphantom +}, \tau_2^+).
\end{equation}
Furthermore, the second-order response functions can be decomposed into three different
parts \cite{chi2-paper}, see Fig.~\ref{fig:chi2-decomposition}:
\begin{equation}
    \rf[2] = \rf[2]_0 + \rf[2]_1 + \vrf[2].
\end{equation}
Here $\rf[2]_0$ is the bare or bubble part that contains diagrams with only one-particle
Green's functions, $\rf[2]_1$ are the first-order terms, which are three-particle diagrams
with a single two-particle vertex $\fv$, and $\vrf[2]$ is the contribution from the full
three-particle vertex $\fv[3]$.

Let us finally also define the Fourier transform,
\begin{equation}
    \rf_{\sigma_1 \sigma_2 \sigma_3}^{\omega_1 \omega_2} =
        \int_0^\beta \! \int_0^\beta \! \rf_{\sigma_1 \sigma_2 \sigma_3}(\tau_1,\tau_2)
        \ee^{\ii \omega_1 \tau_1 + \ii \omega_2 \tau_2} \dl{\tau_1} \dl{\tau_2},
    \label{eq:x3-omega}
\end{equation}
where $\omega_1$ and $\omega_2$ are bosonic Matsubara frequencies.

%% file: three_particle_ladder.tex
\section{Three-particle ladder}
\label{sec:3p-ladder}

The full three-particle vertex is  a complex object, which is expensive
to compute and even storing it becomes difficult if we do not restrict ourselves to its
local part. Fortunately, we can build it completely from simpler two-particle vertices
since the only interaction term we consider is a two-particle interaction. 

In this Section we first express the three-particle vertex as a  a geometric ladder series in terms of irreducible two-particle vertices. 
Our
motivation for this are the ladder diagrams generated by the Bethe--Salpeter equations on
the two-particle level. However calculating the fully irreducible 
vertex in a given three-particle channel is presently not feasible. 

We thus approximate 
the irreducible three particle vertex by a sum of irreducible two-particle vertices.
This makes an actual calculation possible.
If we instead constructed the ladder series in terms of the bare interaction $U$, we would have a three-particle analog of the random phase approximation (RPA). However,
expressing the ladder in terms of the two-particle vertex includes many more Feynman diagrams, among others screening effects.

The local two-particle vertex can be calculated by solving an AIM numerically. The difference to QCD calculations 
with  rainbow-ladder truncation
\cite{Sanchis2013} is that the QCD {calculations are{, to the best of our knowledge,} restricted to one ladder of three quarks, whereas we
calculate the ladder in various pph channels and the ppp channel. Further,} we keep the full frequency dependence of the irreducible two-particle vertex  
that arises from more complicated local Feynman diagrams, rather than just 
the bare effective interaction used in the rainbow-ladder truncation.

\subsection{Three-particle Bethe--Salpeter-like equations}
\label{ssec:3p-bse}

Let us start by writing down the two-particle Bethe-Salpeter equations that we introduced
in \cref{eq:bse} in a very simplified form, 
\begin{equation}
    \fv = \iv_r + \iv_r \cdot \fv,
    \label{eq:simple-bse}
\end{equation}
where $\fv$ is the full vertex, $\iv_r$ are the vertices irreducible in a single channel
$r$, and $\cdot$ denotes the proper connection of two vertex diagrams with Green's
function lines in the correct channel. Generalizing this from the two-particle to the
three-particle level seems straightforward. We define a three-particle diagram as three-particle reducible (3PR) if
it can be separated into two disconnected {three-particle} parts by cutting three Green's function lines,
and as three-particle irreducible (3PI) otherwise.
{Note that a less stringent definition of cutting it into {\em any} kind of parts is not helpful since
then any diagram would be 3PR: it could be simply cut around an outermost two-particle interaction (that we consider here) into two disconnected parts (a 4-particle and a 2-particle one)}.
In \cref{sec:3p-greens-function} we
already identified the ten three-particle channels, so we need to define $\iv[\textrm{3PI}]_{r}$
as the vertex that is 3PI in one {three-particle channel $r$},  use three instead of two Green's functions to
connect everything, and we are done.

\begin{figure}
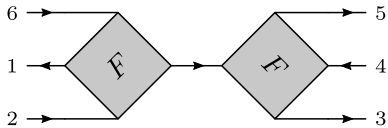

    \centering
    \includeinkscapefigure{Fgf}
    \caption{The only diagram one-particle reducible (1PR) in the channel that separates
             $(612)$ from $(345)$. Looking at \cref{tab:3p-notations} this is the same as
             the three-particle channel $41$.}
    \label{fig:fgf}
\end{figure}

As it turns out, it is not that easy. The two-particle level has the very nice advantage
that all diagrams are inherently one-particle irreducible (1PI) (see
\cref{sec:theoretical-background}). This is not true for the three-particle level. With a
four-point interaction we can easily draw a diagram that goes from three lines down to a
single one and then back to three again. Such a diagram is shown in \cref{fig:fgf}.
Cutting the single Green's function line in the middle separates the diagram into two
connected parts: $(612)$ and $(345)$. This is the same separation that happens when
cutting the three Green's function lines of a diagram 3PR in the $41 = (345)(612)$ channel
(see \cref{tab:3p-notations}). We can do this for all other 
{$\pph$}  channels, too. This shows that
the channels where we encounter one-particle reducibility are the very same as the nine
$\pph$ channels for three-particle reducibility.\footnote{The $\ppp$ channel has no
one-particle equivalent since this would require a two-particle vertex with at least three
incoming lines which violates particle conservation.}

If we were to connect two diagrams that are one-particle reducible (1PR) in the same
channel we would have built a one-particle insertion. Since we are drawing our diagrams
with full Green's functions this would lead to double counting and is therefore not
allowed. To avoid this we do not build a three-particle Bethe--Salpeter-like equation with
$\fv[3]$ and $\iv[\mathrm{3PI}]_{r}$, but subtract diagrams 1PR in $r$ first. We denote the
three-particle vertex that is 1PI in channel $r$ with $\iv_{\mathrm{1PI}, r}$; 
and the one that is 1PI and 3PI in $r$ with $\iv_{\mathrm{(1,3)PI}, r}$. 

Let us note here that one needs to be careful with the 1PR diagrams when calculating physical properties from these. This often involves connecting some of the legs of the three-particle vertex with Green's functions. Depending on the specific pph channel and connection, the 1PR diagrams may (or may not) generate diagrams that are (are not) already contained as a self energy inclusion. Then these must (must not) be discarded.
In our case of the second-order response studied below, this is not the case, the 1PR diagrams contribute.

\begin{figure}
    \centering
    \includeinkscapefigure[width=8cm]{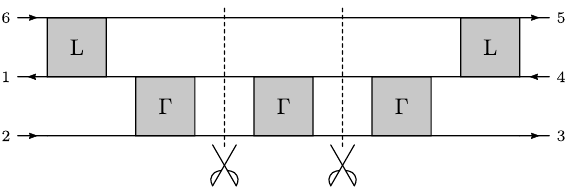}
    \caption{A simple example diagram that illustrates the ambiguity that arises when
             cutting 3PR diagrams into 3PI parts. In our notation  it is reducible in the
             $(345)(612)$ or $41$ channel. Note that cutting after the leftmost or before
             the rightmost vertex is not allowed since then we would end up with
             disconnected parts that are not three-particle vertices anymore. Also, the
             upper vertices are upside down because, 
             in our notation, the orientation of the letter $\Gamma$ is with respect to the incoming and outgoing particles as defined  in
             \cref{fig:g2-frequency-notations} for the standard orientation.}
    \label{fig:ambiguous-3pr-cuts}
\end{figure}

\begin{figure*}
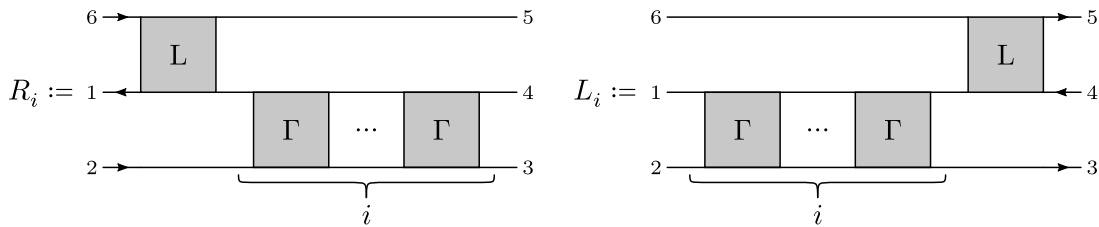

    \centering
    \includeinkscapefigure{2prDoubleCounting}
    \caption{Definition of $R_i$ and $L_i$. Again, the upper vertices are upside down
             because of how we defined them in \cref{fig:g2-frequency-notations}.}
    \label{fig:2pr-double-counting}
\end{figure*}

Even with the $\iv_{\mathrm{1PI}, r}$ and $\iv_{\mathrm{(1,3)PI}, r}$ thus defined, there is still one remaining problem: some 3PR
diagrams cannot be uniquely cut into 3PI parts. A simple example of this is shown in
\cref{fig:ambiguous-3pr-cuts}, where there are two possible ways to cut the diagram and
each one results in different 3PI parts.
The underlying issue is that 3PI diagrams can be two-particle reducible (2PR) in related
channels. This means that two-particle vertices can be added or cut off from those
diagrams without changing their three-particle reducibility. If we use the notation
defined in \cref{fig:2pr-double-counting}, where $R_i$ has $i$ two-particle vertices on
the right and $L_i$ has $i$ vertices on the left, we see that
\begin{equation}
    R_i \cdot L_j = R_n \cdot L_m, \quad \forall \ i + j = n + m.
    \label{eq:RdotL}
\end{equation}
Since both $R_i$ and $L_i$ are elements of $\iv_{\mathrm{1PI}, r}$ and $\iv_{\mathrm{(1,3)PI}, r}$ we would again
invoke double counting.

One way to prevent this and make cutting off 3PI parts unique is to require that we always
cut off the \enquote{leftmost} part, i.e., 
{that we fix $m=i+j$ and $n=0$ in \cref{eq:RdotL}} \footnote{This is equivalent  to always include as many diagrammatic components as possible in the reducible (right) part of the vertex.} This is equivalent to requiring that the vertices 1PI
and 3PI in a channel $r$ must also be two-particle irreducible (2PI) in the three related
two-particle channels \enquote{on the right}. Denoting this quantity with $\iv_{(1,\vec{2},3)\,\mathrm{PI}, r}$ 
we can finally write down the three-particle equivalent of the
Bethe--Salpeter equation:
\begin{equation}
    \iv_{1\mathrm{PI}, r} = \iv_{(1,\vec{2},3)\,\mathrm{PI}, r}  + \iv_{(1,\vec{2},3)\,\mathrm{PI}, r}\cdot \iv_{1\mathrm{PI}, r}.
    \label{eq:3p-bse}
\end{equation}

To recover the full vertex $\fv[3]$ we need to add the diagrams 1PR in channel $r$. As it
turns out there is only a single one for each channel. They are all of the same form: two
full two-particle vertices $\fv[2]$ connected by a single Green's function line $\gf$.
One of them is depicted in \cref{fig:fgf}. With this we have
\begin{equation}
    \fv[3] = \iv_{\mathrm{1PI}, r} + (\fv[2] \gf \fv[2])_r.
    \label{eq:f3-gamma1-fgf}
\end{equation}
For a more detailed discussion on three-particle (ir)reducibilities see the Supplemental Material~\cite{SM}.

Let us also note here that our Bethe-Salpeter-like equations are similar to the Faddeev equations employed in other fields of physics.
The fundamental difference is that Faddeev~\cite{Faddeev1961} considered a three-particle
Hamiltonian whereas we consider three-particles and holes that propagate through a solid. Thereby they strongly interact with many other  electrons. Consequently, in-between the three incoming and outgoing particles(holes), there can be further particle(hole) pairs generated through the Coulomb interaction. This is hidden here in the more complicated Feynman diagrams of the irreducible vertex $\iv_{(1,\vec{2},3)\mathrm{PI}, r}${.} The opposite effect that we can have, a particle-hole annihilation in-between, is more apparent in the equations, and leads to the additional contribution of the 1PR diagram of \cref{fig:fgf}.
This contribution is absent in the ppp channel and the Faddeev equations.

\subsection{Approximate three-particle ladder}
\label{ssec:approximate-3p-ladder}

In the last section we saw that the exact three-particle ladders generated by the
Bethe--Salpeter-like equations are built from three-particle vertices that are 1PI and 3PI
in a certain channel and 2PI in the three related channels \enquote{on the right}. The
problem is that we have no easy way to compute them. We could construct them order by order
from two-particle vertices, but that is 
tedious and unfeasible.
Instead, we propose approximating the exact three-particle ladder with a ladder of
two-particle vertices.

\begin{figure}
    \centering
    \includeinkscapefigure[width=8cm]{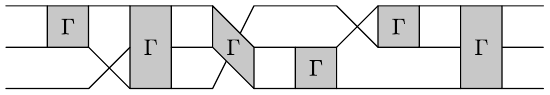}
    \caption{Example diagram for an approximate three-particle ladder built only from
             two-particle vertices. Only Green's function lines touching the corners of a
             vertex are really connected to it. The middle line on the right side, e.g.,
             is not connected to the right-most vertex but runs through \enquote{behind}
             it.}
    \label{fig:example-ladder}
\end{figure}

\begin{figure*}
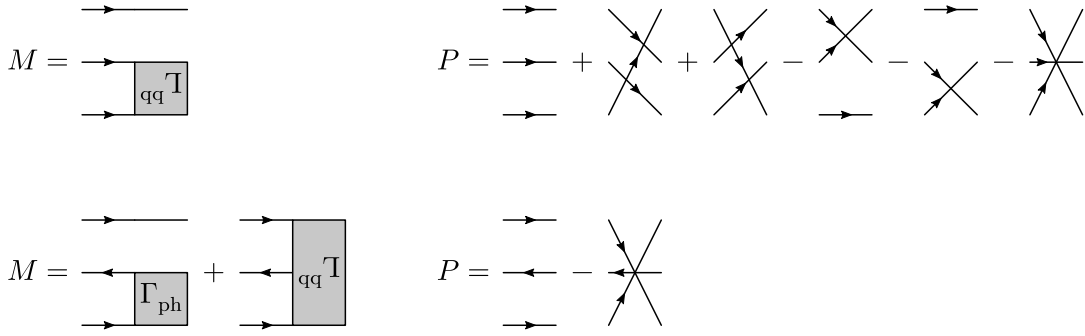

    \centering
    \includeinkscapefigure{MAndP}
    \caption{Diagrammatic representation of $M$ and $P$ for $\ppp$ ladders (top) and
             $\pph$ ladders (bottom). The $\pp$ vertex is mirrored because the frequencies of the incoming and outgoing lines are opposite to the definition  in \cref{fig:g2-frequency-notations}.}
    \label{fig:m-and-p}
\end{figure*}

The general idea is to find all possible ways to connect a two-particle vertex to three
lines on the left and three lines on the right, and then build a ladder from that.
\Cref{fig:example-ladder} shows an example diagram for such a ladder.
Formally, we can write this as
\begin{equation}
    \tilde{L} = \sum_{n = 1}^\infty P(M P)^n,
    \label{eq:naive-ladder}
\end{equation}
where $\tilde{L}$ is the approximate ladder, $P$ consists of all permutations of the three
lines of the ladder and $M$ contains a two-particle vertex.  Since the ladder can have
multiple vertices right next to each other, using the full two-particle vertex in $M$
would lead to double counting. To avoid this issue we use $\iv_r$, the two-particle vertex
irreducible in channel $r$. Which channel we need depends on which kind of
three-particle ladder we build: either one with three particles ($\ppp$) or one with two
particles and a hole ($\pph$).\footnote{The two cases with three holes or two holes and
one particle can easily be obtained from the other two via a particle--hole
transformation.} In the $\ppp$ case we only need $\iv_{\pp}$, in the $\pph$ case we need
to add $\iv_{\ph}$ as well. The permutation matrix $P$ also differs between the two cases
because we are not allowed to swap particles with holes and vice versa. Thus, for a $\pph$
ladder the permutation matrix only consists of two instead of $3! = 6$ terms.
\Cref{fig:m-and-p} shows a diagrammatic representation of $M$ and $P$ for both cases.

From now on we only focus on the $\pph$ ladder since it directly matches with our
definition of the three-particle Green's function shown in \cref{fig:g3}. Because of that
it is also easier to compute the vertex corrections to the second-order response functions
$\vrf[2]$ from it, which is convenient for our comparison tests. It is, however,
straightforward to apply the ideas, concepts, issues, and solutions presented in the rest
of this section to the $\ppp$ case.

Let us continue by introducing some notation. We label the lines of the ladder with the
numbers $1$ to $3$ starting from the top. Since in the first term of $M$ shown in
\cref{fig:m-and-p} (bottom left) the $\ph$ vertex sits between lines $2$ and $3$ we call
it $M_{23}$. The second term with the $\pp$ vertex is then called $M_{13}$, and we can
write
\begin{equation}
    M = M_{23} + M_{13}.
    \label{eq:m-m-23-m13}
\end{equation}
In combination with the permutation matrix $P$ the crossing symmetry of $M_{13}$,
inherited from $\iv_{\pp}$, causes the following double counting issue:
\begin{equation}
    P M_{13} = M_{13} P = 2 M_{13}.
    \label{eq:m13-double-counting}
\end{equation}
This can easily be corrected with an additional factor of $1/2$.

\begin{figure}
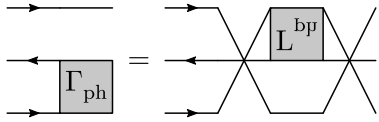

    \centering
    \includeinkscapefigure{M23M12}
    \caption{Diagrammatic representation of \cref{eq:m23-m12}, the relation between
             $M_{12}$ and $M_{23}$}
    \label{fig:m23-m12}
\end{figure}

\begin{figure}
    \centering
    \includeinkscapefigure[width=8cm]{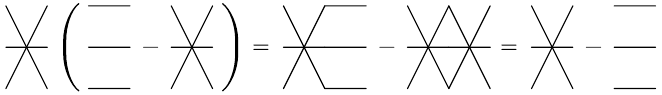}
    \caption{Diagrammatic representation of \cref{eq:p13-p}}
    \label{fig:p13-p}
\end{figure}

\Cref{eq:m13-double-counting} also shows that $M_{13}$ commutes with $P$. If this were
true for the whole $M$ we could simplify \cref{eq:naive-ladder}, by pulling all
permutation matrices to one side, disentangling the ladder. $M_{23}$ prevents this because
swapping the two particle lines folds the $\ph$ vertex up, creating a new diagram, which we
call $M_{12}$ (it connects the first and second line), that is not in $M$. However, swapping the particle lines of $M_{12}$
brings us back to $M_{23}$, so the two diagrams turn into each other under the considered
permutation. This means that in order to make $M$ commute with $P${,} all we need to do is to
symmetrize its $\ph$ part and write it as the sum of both diagrams. Denoting the
permutation matrix that exchanges the two particle lines with $P_{13}$, we can write
$M_{23}$ in terms of $M_{12}$:
\begin{equation}
    M_{23} = P_{13} M_{12} P_{13}.
    \label{eq:m23-m12}
\end{equation}
\Cref{fig:m23-m12} shows a diagrammatic representation of this.
Applying $P_{13}$ to the whole permutation matrix $P$ only gives a sign:
\begin{equation}
    P_{13} P = P P_{13} = P_{13} - P_{13} P_{13} = P_{13} - \mathbb{I} = -P,
    \label{eq:p13-p}
\end{equation}
where $\mathbb{I}$ is the identity permutation. The diagrams for this equation can be found in
\cref{fig:p13-p}.
Putting everything together we can compute the desired symmetrization,
\begin{equation}
    P M_{23} P = \frac{1}{2} P (M_{23} + P_{13} M_{12} P_{13}) P
        = \frac{1}{2} P (M_{23} + M_{12}) P.
    \label{eq:m23-symmetrized}
\end{equation}
For later convenience we pull out the factors of $1/2$ from \cref{eq:m23-symmetrized} and
the double counting issue in \cref{eq:m13-double-counting} when redefining $M$:
\begin{equation}
    M = M_{12} + M_{23} + M_{13}.
\end{equation}
This new $M$ commutes with $P \,$:
\begin{eqnarray}
    P M &=& M - P_{13} M_{12} - P_{13} M_{23} - P_{13} M_{13}
        \nonumber \\ 
        &=& M - M_{23} P_{13} - M_{12} P_{13} - M_{13} P_{13}
        = M P.
\end{eqnarray}
With the previously pulled out factor of $1/2$, we get
\begin{equation}
    \tilde{L} = \sum_{n = 1}^\infty P \, \left( \frac{1}{2} M P \right)^n
              = \sum_{n = 1}^\infty M^n P \, \left( \frac{1}{2} P \right)^n
\end{equation}
for the approximate ladder. One can easily check that $P/2$ is idempotent, and we can thus
further simplify the ladder equation:
\begin{equation}
    \tilde{L} = \sum_{n = 1}^\infty M^n P \, \left( \frac{1}{2} P \right)^n
              = \sum_{n = 1}^\infty M^n P \, \frac{1}{2} P
              = \sum_{n = 1}^\infty M^n P.
    \label{eq:disentangled-ladder}
\end{equation}
The ladder is now disentangled, with only a single permutation matrix at one end. This
also shows that the ladder is crossing symmetric with respect to the two particle lines: it
fulfills crossing symmetries with respect to exchange of incoming particle legs on the left, and of outgoing ones at the right. The full three-particle vertex, however, is crossing
symmetric with respect to all particle and hole lines: it has six crossing symmetries (four with respect to exchanging particle or hole legs from left to right).
This difference in symmetries is an important issue that we will come back to later.

The simpler form of \cref{eq:disentangled-ladder} also reveals that we generate a lot of
disconnected terms. They are the ones where we do not mix the different components of $M$
and therefore build a two-particle ladder with a disconnected Green's function running
next to it. There is one such disconnected term for each ladder. Subtracting them leaves
us with
\begin{equation}
    L = \sum_{n = 1}^\infty (M^n - M_{12}^n - M_{23}^n - M_{13}^n) P.
    \label{eq:connected-ladder}
\end{equation}

Now let us focus on the last double counting issue. Again, it comes from the irreducible
$\pp$ vertex in $M_{13}$, but this time it has nothing to do with the permutation matrix.
$\iv_{\pp}$ is the sole perpetrator and this issue already appears on the two-particle
level. There it mandates the factor of $1/2$ in the Bethe--Salpeter equation of the $\pp$
channel [see \cref{eq:bse}].

\begin{figure*}
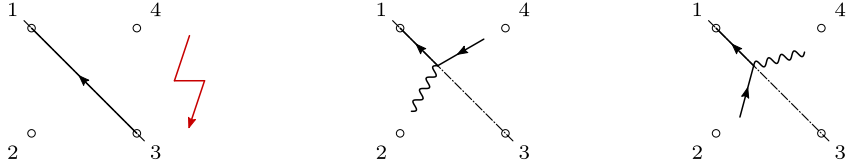

    \centering
    \includeinkscapefigure{ImpossibleCrossingSymmetry}
    \caption{Visual proof that a diagram with a two-particle interaction cannot be
             symmetric along a diagonal. The first diagram is not allowed since in our
             convention an annihilation operator sits at both coordinates "1" and
             "3"; see \cref{eq:g2-tau}. The second and third diagram
             illustrate that it is impossible to add an interaction term and keep the
             diagram symmetric along the diagonal. Therefore, a single diagram cannot be
             crossing symmetric on its own.}
    \label{fig:impossible-crossing-symmetry}
\end{figure*}

Let us hence start by looking at the crossing symmetry for a two-particle vertex diagram.
With a two-particle interaction term it is not possible that any given diagram
contributing to the vertex is crossing symmetric on its own. The reason is that with our
established conventions such a diagram would have to be symmetric along one of the two
diagonals ($1$--$3$ or $2$--$4$) and that is impossible as illustrated in
\cref{fig:impossible-crossing-symmetry}.
Without any interaction terms we would have to draw a Green's function line from $1$ to
$3$, which is not allowed because in our convention an annihilation operator sits at both
coordinates [see \cref{eq:g2-tau}]. Adding an interaction term effectively splits
the Green's function line into an interaction line and another Green's function line. No
matter at what angle or on what side we draw that this will never be symmetric along the
$1$--$3$ diagonal because they are two different types of lines.

Objects like the two-particle Green's function, full vertex, or irreducible $\pp$ vertex
are crossing symmetric because they consist of a sum of terms and this sum contains pairs
of diagrams that turn into each other under crossing exchange. The reason for the diagrams
to come in pairs and not in larger groups is that crossing is an involution, i.e.,
applying it twice yields the identity. Probably the simplest example for such a pair of
diagrams are the disconnected parts of the two-particle Green's function $\gf[2]$ shown in
\cref{fig:g2-decomposition}. Swapping $1 \leftrightarrow 3$ or $2 \leftrightarrow 4$ turns
the first diagram into the second and vice versa. For a general discussion let $\gamma_1,
\gamma_2 \in \iv_{\pp}$ be a crossing-related pair of diagrams, i.e.,
\begin{equation}
    \begin{split}
        \gamma_1 & = C \gamma_2 = \gamma_2 C \\
        \gamma_2 & = C \gamma_1 = \gamma_1 C
    \end{split}
\end{equation}
where $C$ denotes the crossing operation. If two irreducible $\pp$ vertices are connected
with two particle lines both terms $\gamma_1 \gamma_2$ and $\gamma_2 \gamma_1$ appear.
However, since crossing is its own inverse we can write
\begin{equation}
    \gamma_1 \gamma_2 = \gamma_1 C C \gamma_2 = \gamma_2 \gamma_1.
\end{equation}
Therefore, we count all diagrams twice and need the factor of $1/2$ in the Bethe--Salpeter
equation for the $\pp$ channel.

For the three-particle ladder the case is similar. The main difference is that we can have
diagrams from $\ph$ vertices in between the $\pp$ ones, so we need to consider that in our
argument. Let $m$ be a general combination of $\ph$ diagrams that can appear in the
ladder, i.e., $m \in (M_{12} + M_{23})^n$. Its crossing-related counterpart $\bar{m} =
P_{13} m P_{13}$ is the same diagram but with the $M_{12}$ and $M_{23}$ parts swapped
[remember \cref{eq:m23-m12}], so it is also generated by the ladder. If we pick an
arbitrary pair of crossing-related diagrams $\gamma_1, \gamma_2 \in M_{13}$, the ladder
generates both $\gamma_1 m \gamma_2$ and $\gamma_2 \bar{m} \gamma_1$. Using the same trick
as before we can write
\begin{equation}
    \gamma_1 m \gamma_2
        = \gamma_1 P_{13} P_{13} m P_{13} P_{13} \gamma_2
        = \gamma_2 \bar{m} \gamma_1,
\end{equation}
and show that we count all such diagrams twice. Therefore, we must add a factor of $1/2$
in front of $\iv_{\pp}$, as well.

\begin{figure}
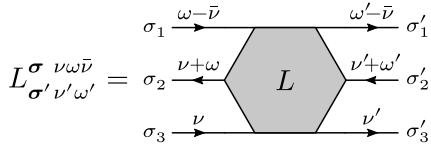

    \centering
    \includeinkscapefigure{LadderNotation}
    \caption{Frequency and spin notation for the three-particle ladder}
    \label{fig:ladder-notation}
\end{figure}

\begin{figure*}
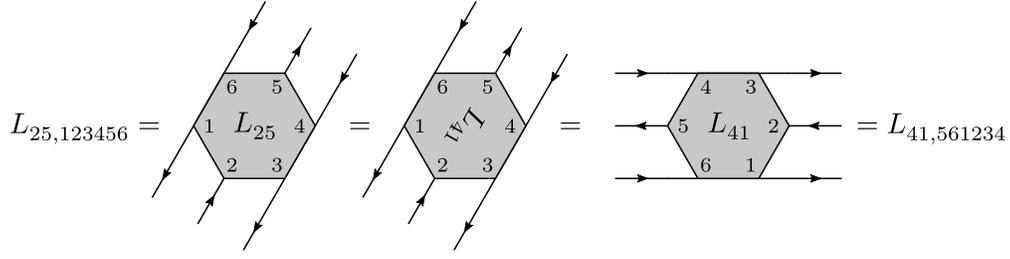

    \centering
    \includeinkscapefigure{LadderConversion}
    \caption{The $\pph$ ladder in the $25$ channel, obtained from the one in the
             $41$ channel}
    \label{fig:ladder-conversion}
\end{figure*}

With the last topological issue resolved we can finally work out the full equations for
the three-particle ladder and its components with all prefactors, indices, and arguments.
So far we always drew horizontal ladders in the $41$ channel, so we choose the matching
frequency notation. Both the channel and its frequency notation are defined in
\cref{tab:3p-notations}. A diagrammatic representation with all spin and frequency indices
is shown in \cref{fig:ladder-notation}.
Ladders in different $\pph$ channels can easily be obtained by rotation or swapping of
external legs. An example of this is illustrated in \cref{fig:ladder-conversion}.
Due to implementation details the frequency notations for the irreducible two-particle
vertices are not the ones introduced in
\cref{tab:g2-frequency-notations,fig:g2-frequency-notations}. Instead, all frequencies are
shifted by $\omega$ as shown in \cref{fig:gamma-notations}.
\begin{figure*}
    \centering
    \includeinkscapefigure{GammaNotations}
    \caption{Frequency and spin notations for the two-particle vertices used in the
             three-particle ladder}
    \label{fig:gamma-notations}
\end{figure*}
We also denote frequency and spin indices on the right side of the diagrams as subscripts
to save horizontal space.

To connect the two-particle vertices in the ladder we need Green's functions. We choose to
add two on the right of each vertex in the $M$'s and then three on the very left of the
entire ladder. This means that the ladder is not an approximation for a \enquote{naked}
three-particle vertex but has Green's functions attached to every external point. This is
convenient since we need them anyway for computing the vertex contribution to second-order
response functions $\vrf[2]$. Putting everything together yields
\begin{align}
    M_{12, \vphantom{\sigma'_3}}^{\vphantom{\sigma_3}}
        {}^{\vect{\sigma} \vphantom{\sigma_3}}_{\vect{\sigma}' \vphantom{\sigma'_3}}
        {}^{\nu \omega \nubar \vphantom{\sigma_3}}_{\nu' \omega' \vphantom{\sigma'_3}}
        & \Def \frac{1}{\invtemp}
        \delta^{\sigma_3}_{\sigma'_3}
        \delta^{\nu \vphantom{\sigma_3}}_{\nu' \vphantom{\sigma'_3}}
        \iv_{\ph, \vphantom{\sigma'_3}}^{\vphantom{\sigma_3}}
            {}^{\sigma_2 \sigma_1}_{\sigma'_2 \sigma'_1}
            {}^{(\omega - \nubar) (\nu + \nubar)}_{(\omega' - \nubar)}
        \gf^{\omega' - \nubar \vphantom{\sigma_3}}_{\vphantom{\sigma'_3}}
        \gf^{\omega' + \nu \vphantom{\sigma_3}}_{\vphantom{\sigma'_3}},
    \label{eq:m12}
    \\
    M_{23, \vphantom{\sigma'_3}}^{\vphantom{\sigma_3}}
        {}^{\vect{\sigma} \vphantom{\sigma_3}}_{\vect{\sigma}' \vphantom{\sigma'_3}}
        {}^{\nu \omega \nubar \vphantom{\sigma_3}}_{\nu' \omega' \vphantom{\sigma'_3}}
        & \Def \frac{1}{\invtemp}
        \delta^{\sigma_1}_{\sigma'_1}
        \delta^{\omega \vphantom{\sigma_3}}_{\omega' \vphantom{\sigma'_3}}
        \iv_{\ph, \vphantom{\sigma'_3}}^{\vphantom{\sigma_3}}
            {}^{\sigma_2 \sigma_3}_{\sigma'_2 \sigma'_3}
            {}^{\nu \omega \vphantom{\nu' \sigma_3}}_{\nu' \vphantom{\sigma'_3}}
        \gf^{\nu'\vphantom{\sigma_3}}_{\vphantom{\sigma'_3}}
        \gf^{\omega + \nu' \vphantom{\sigma_3}}_{\vphantom{\sigma'_3}},
    \label{eq:m23}
    \\
    M_{13, \vphantom{\sigma'_3}}^{\vphantom{\sigma'_3}}
        {}^{\vect{\sigma} \vphantom{\sigma'_3}}_{\vect{\sigma}' \vphantom{\sigma'_3}}
        {}^{\nu \omega \nubar \vphantom{\sigma'_3}}_{
            \nu' \omega' \vphantom{\sigma'_3}}
        & \Def \frac{1}{2 \invtemp}
        \delta^{\sigma_2}_{\sigma'_2}
        \delta^{\nu + \omega \vphantom{\sigma'_3}}_{
            \nu' + \omega' \vphantom{\sigma'_3}}
        \iv_{\pp, \vphantom{\sigma'_3}}^{\vphantom{\sigma'_3}}
            {}^{\sigma'_1 \sigma'_3}_{\sigma_1 \sigma_3 \vphantom{\sigma'_3}}
            {}^{\nu (\omega + \nu - \nubar)}_{\nu' \vphantom{\sigma'_3}}
        \gf^{\omega' - \nubar \vphantom{\sigma'_3}}_{\vphantom{\sigma'_3}}
        \gf^{\nu'\vphantom{\sigma'_3}}_{\vphantom{\sigma'_3}},
    \label{eq:m13}
\end{align}
for the components of $M$, where $\vect{\sigma} = (\sigma_1, \sigma_2, \sigma_3)$ and
similarly for $\vect{\sigma}'$. The factors $1 / \invtemp$ ensure that $M$ is
dimensionless, which is necessary because we sum over different powers of them in the
ladder. A diagrammatic representation of the $M$'s is shown in \cref{fig:ms}.
\begin{figure}
    \centering
    \includeinkscapefigure{Ms}
    \caption{Diagrammatic representation of \cref{eq:m12,eq:m23,eq:m13}, the definitions
             of the three components of $M$}
    \label{fig:ms}
\end{figure}
The permutation matrix is simply given by a combination of Kronecker deltas
\begin{equation}
    P^{\vect{\sigma} \vphantom{\sigma_1}}_{\vect{\sigma}' \vphantom{\sigma'_1}}
        {}^{\nu \omega \nubar \vphantom{\sigma_1}}_{\nu' \omega' \vphantom{\sigma'_1}}
        = \delta^{\sigma_1}_{\sigma'_1}
        \delta^{\sigma_2}_{\sigma'_2}
        \delta^{\sigma_3}_{\sigma'_3}
        \delta^{\omega \vphantom{\sigma_1}}_{\omega' \vphantom{\sigma'_1}}
        \delta^{\nu \vphantom{\sigma_1}}_{\nu' \vphantom{\sigma'_1}}
        - \delta^{\sigma_1}_{\sigma'_3}
        \delta^{\sigma_2}_{\sigma'_2}
        \delta^{\sigma_3}_{\sigma'_1}
        \delta^{\omega - \nubar \vphantom{\sigma_1}}_{\nu' \vphantom{\sigma'_1}}
        \delta^{\nu + \omega \vphantom{\sigma_1}}_{\nu' + \omega' \vphantom{\sigma'_1}}.
    \label{eq:p}
\end{equation}
Finally the full equation for the approximate three-particle ladder reads
\begin{eqnarray}
    L^{\vect{\sigma} \vphantom{\sigma_1}}_{\vect{\sigma}' \vphantom{\sigma'_1}}
        {}^{\nu \omega \nubar \vphantom{\sigma_1}}_{\nu' \omega' \vphantom{\sigma'_1}}
        &=& -\beta^2
        G^{\omega - \nubar \vphantom{\sigma_1}}_{\vphantom{\sigma'_1}}
        G^{\nu + \omega \vphantom{\sigma_1}}_{\vphantom{\sigma'_1}}
        G^{\nu \vphantom{\sigma_1}}_{\vphantom{\sigma'_1}}
        \!\!\! \sum_{\sigma_1, \nu_1, \omega_1} \!\! \Big(\sum_{n = 0}^{\infty} \big(
            M^n - M_{12}^n
            \nonumber \\ && \phantom{-\beta}
            - M_{23}^n - M_{13}^n \big) + 2 \Big)
            {}^{\! \vect{\sigma} \vphantom{\sigma_1}}_{
                \! \vect{\sigma}_1 \vphantom{\sigma'_1}}
            {}^{\nu \omega \nubar \vphantom{\sigma_1}}_{
                \nu_1 \omega_1 \vphantom{\sigma'_1}}
        P^{\vect{\sigma}_1}_{\vect{\sigma}' \vphantom{\sigma'_1}}
            {}^{\nu_1 \omega_1 \nubar}_{\nu' \omega' \vphantom{\sigma'_1}},
    \label{eq:l}
\end{eqnarray}
where the prefactor of $\invtemp^2$ ensures a dimension of $\dimension{\tau}^5$, as it
should be for a quantity depending on five Matsubara frequencies. The addition of $2$ is
necessary because the sum now starts at $n = 0$ to turn it into a proper geometric series.
The notation also shows that the equation is diagonal in $\nubar$ and the terms can be
viewed as simple matrices when condensing $(\vect{\sigma}, \nu, \omega)$ into a single
compound index. This makes it easy to employ the closed-form formula for the geometric
series.

Substituting the 1PI vertex, $\iv_{\mathrm{1PI}, r}$, in \cref{eq:f3-gamma1-fgf}, with the approximate
three-particle ladder, $L$, turns out to yield even qualitatively bad results. The reason
for that lies in the crossing symmetries. As mentioned before, the approximate ladder only
satisfies two of the six crossing symmetries of the full three-particle vertex $\fv[3]$.
The 1PR diagram $(\fv[2] \gf \fv[2])_r$ in \cref{eq:f3-gamma1-fgf} is in the same channel
$r$ as $\iv_{\mathrm{1PI}, r}$ or in this case the ladder, and therefore has the same problem. To
solve this we compute the average of the approximate ladder and the 1PR diagram $(\fv[2]
\gf \fv[2])_r$ over all nine $\pph$ channels defined in \cref{tab:3p-notations}. The full
three-particle vertex (with legs) in this approximation is thus given by
\begin{equation}
    \gf \gf \gf \fv[3] \gf \gf \gf
        \approx \frac{1}{9} \sum_{r}[ L_{r}
        + \gf \gf \gf (\fv[2] \gf \fv[2])_{r} \gf \gf \gf],
    \label{eq:f3-ladder}
\end{equation}
where $r$ runs over the nine $\pph$ channels.

Averaging is preferable to summation to avoid overcounting of lower-order diagrams. 
{Considering second-order diagrams of $G_3$, every $\pph$ ladder generates 8 (out of a total of 9 of them)} and the 1PR diagram added separately is the
9th. This means that a single channel already contains all diagrams up to second order
in $\fv[2]$ and not averaging would lead to overcounting them by a factor of nine.
Averaging guarantees  that the thus calculated vertex is correct to second order
\footnote{Note, ppp channel contributions need not be considered extra. They  are identical to particle-hole-symmetry related pph ladder diagrams.}.


Note that we do not need to come up with nine versions of \cref{eq:l} to compute the
ladders in all $\pph$ channels. Instead, it is much easier to convert the ladder in the
$41$ channel to the other ones by rotation or swapping of legs as exemplified in
\cref{fig:ladder-conversion}.

\subsection{Second-order response function}
\label{ssec:2ndorderchi}

The last step in our computations is using the approximate three-particle vertex
and calculating its contribution to the second-order response function $\vrf[2]$, which we
here compare to the exact solution. Since our approximation already comes attached with
six Green's functions, this is, at least in theory, easily done by summing over the
fermionic frequencies:
\begin{equation}
    \vrf^{\omega \omega'}
        = \sum_{\nu \nu' \nubar}
          (\gf\gf\gf \fv[3] \gf\gf\gf)^{\nu \omega \nubar}_{\nu' \omega'}.
    \label{eq:chi2-from-vertex}
\end{equation}
In numerical calculations these infinite sums are of course limited by the grid size. As
it turns out it is important how exactly the sums are implemented because some terms need
to cancel. The reason for that is that the imaginary part of $\vrf_{\uuu}$ and
$\vrf_{\uud}$ must vanish. This can easily be shown when remembering that complex
conjugation inverts the order of spins and fermionic frequencies while also negating the
latter \cite{Rohringer2013a}:\footnote{The cited reference only shows that $\gf[3]$
behaves like that under complex conjugation. With a straightforward although tedious
calculation one can show, however, that the six, fully disconnected parts of $\gf[3]$ and
the nine partially connected parts follow the same equations. Therefore, the full vertex
$\fv[3]$ must follow them as well.}
\begin{equation}
    \big(\fv_{123456}^{\nu_1 \dots \nu_6}\big)^* = \fv_{654321}^{-\nu_6 \dots -\nu_1}.
    \label{eq:f3-complex-conjugation}
\end{equation}
If we now use, e.g., the first $\ph$ notation from \cref{tab:g3-2p-frequency-notations},
\begin{equation}
    \begin{split}
        \omega_a & = \nu_1 - \nu_2, \\
        \omega_b & = \nu_3 - \nu_4, \\
        \omega_c & = \nu_5 - \nu_6,
        \label{eq:omegaabc}
    \end{split}
\end{equation}
we see that the bosonic frequencies also change their order but keep their signs after
complex conjugation:
\begin{equation}
    \begin{split}
        \omega_a & {} \to -\nu_6 + \nu_5 = \omega_c, \\
        \omega_b & {} \to -\nu_4 + \nu_3 = \omega_b, \\
        \omega_c & {} \to -\nu_2 + \nu_1 = \omega_a.
    \end{split}
    \label{eq:omega-cc}
\end{equation}
For the second-order response functions we therefore have
\begin{equation}
    \left(\vrf_{123456}^{\omega_a \omega_b \omega_c}\right)^*
        = \vrf_{654321}^{\omega_c \omega_b \omega_a}.
\end{equation}
Using the swapping symmetries introduced in \cite[Eqs. (G3)--(G5)]{chi2-paper}, we see that the
$\uuu$ and $\uud$ components of $\vrf[2]$ are indeed purely real:
\begin{align}
    \left(\vrf_{\uuu}^{\omega_a \omega_b \omega_c}\right)^*
        & = \vrf_{\uuu}^{\omega_c \omega_b \omega_a}
        = \vrf_{\uuu}^{\omega_a \omega_b \omega_c},
    \label{eq:chi2-vertex-1-cc}
    \\
    \left(\vrf_{\uud}^{\omega_a \omega_b \omega_c}\right)^*
    & = \vrf_{\duu}^{\omega_c \omega_b \omega_a}
    = \vrf_{\uud}^{\omega_a \omega_b \omega_c}.
    \label{eq:chi2-vertex-4-cc}
\end{align}

\begin{figure}
    \centering
    \includeinkscapefigure[width=8.8cm]{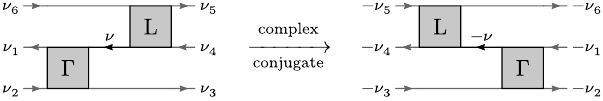}
    \caption{Diagrammatic representation of \cref{eq:cc-ladder-term}. The complex
             conjugation mirrors the positions of the vertices and frequencies along the
             vertical axis. The frequencies are also negated.}
    \label{fig:cc-ladder-term}
\end{figure}

The problem for the approximate ladder is that it does not have those swapping
symmetries.\footnote{The swapping symmetries are just a combination of two crossing
symmetries, and we already pointed out that the ladder is missing most of the latter.}
Fortunately, they are not required, if we do the complex conjugation of the ladder
properly and choose the right $\ph$ notation. To show what we mean with the first part,
let us do an explicit complex conjugation for a term that appears in the ladder. A simple
example is
\begin{equation}
    \big(\iv^{\nu_1 \nu_2 \nu_3 \nu} \, \gf^{\nu} \, \iv^{\nu \nu_6 \nu_5 \nu_4}\big)^*
        \! = \iv^{\nu -\nu_3 -\nu_2 -\nu_1} \, \gf^{-\nu} \,
        \iv^{-\nu_4 -\nu_5 -\nu_6 -\nu}.
    \label{eq:cc-ladder-term}
\end{equation}
The corresponding Feynman diagrams are found in \cref{fig:cc-ladder-term} and show that
the positions of the vertices and frequencies get mirrored along the vertical axis with
the frequencies also being negated.
This holds in general for all ladder terms as can be seen when computing a few more simple
diagrams and then using induction. Since the ladder generates all possible arrangements of
two-particle vertices it also contains all mirrored terms. Therefore, the correct way to
complex conjugate a ladder is
\begin{equation}
    \big(L_{123456}^{\nu_1 \dots \nu_6}\big)^*
        = L_{432165}^{-\nu_4 -\nu_3 -\nu_2 -\nu_1 -\nu_6 -\nu_5}.
\end{equation}
This does not contradict \cref{eq:f3-complex-conjugation} since the full three-particle
vertex has all crossing and swapping symmetries, and we can write
\begin{equation}
    \big(\fv_{123456}^{\nu_1 \dots \nu_6}\big)^*
        = \fv_{654321}^{-\nu_6 \dots -\nu_1}
        = \fv_{432165}^{-\nu_4 -\nu_3 -\nu_2 -\nu_1 -\nu_6 -\nu_5}.
    \label{eq:f3-complex-conjugation-2}
\end{equation}

To ensure that the $\uuu$ and $\uud$ components of $\vrf[2]$ are real, we have to find a
$\ph$ notation where the imaginary parts of the ladder cancel when summing over the
fermionic frequencies. This means that the complex conjugate of a frequency component of
$L$ must map to the same bosonic frequencies. The $\phab$ notation from
\cref{tab:g3-2p-frequency-notations}  achieves exactly that:
\begin{equation}
    \begin{split}
        \omegabar_a & = \nu_1 - \nu_4 \ \to\  -\nu_4 + \nu_1 = \omegabar_a, \\
        \omega'_b   & = \nu_3 - \nu_2 \ \to\  -\nu_2 + \nu_3 = \omega'_b, \\
        \omega_c    & = \nu_5 - \nu_6 \ \to\  -\nu_6 + \nu_5 = \omega_c,
    \end{split}
    \label{eq:phia}
\end{equation}
and therefore
\begin{equation}
    \left(L_{123456}^{\nu_a \omegabar_a \nu_b \omega'_b \nu_c}\right)^*
        = L_{432165}^{-\nu_a \omegabar_a -\nu_b \omega'_b -\nu_c}.
\end{equation}
In the numerical implementation we only have to make sure that we sum over a symmetric
interval of fermionic frequencies to ensure cancellation.

So far we only considered the approximate three-particle ladder in the $41$ channel.
However, in \cref{eq:f3-ladder} we average over ladders in all nine $\pph$ channels.
Fortunately, this does not cause additional issues with the cancellation of imaginary
parts: When converting the ladder in the $41$ channel to some other channel we have to
swap one or two pairs of legs. Channels $23$ and $65$ are the ones requiring two pairs.
Their Feynman diagrams turn out to be symmetric along the vertical axis, so the same
considerations as for the $41$ channel hold, and the imaginary parts cancel. The remaining
six channels are not symmetric on their own, but they form mirror pairs: $25$ and $63$,
$45$ and $61$, as well as $21$ and $43$. When summing over those pairs they cancel each
other's imaginary parts for the $\uuu$ and $\uud$ component as well.

Putting everything together we end up with the following approximation for the vertex
contribution to the second-order response function:
\begin{eqnarray}
    \vrf^{\omegabar_a \omega'_b}
        &\approx& \sum_{\nu_a \nu_b \nu_c} \frac{1}{9} \sum_{r} [
        L^{\nu_a \omegabar_a \nu_b \omega'_b \nu_c}_{r}
       \\ && \nonumber
       + (\gf \gf \gf (\fv[2] \gf \fv[2])_{r} \gf \gf \gf)^{
            \nu_a \omegabar_a \nu_b \omega'_b \nu_c}],
    \label{eq:chi2-vertex-ladder}
\end{eqnarray}
where the ladder $L$ and the 1PR diagram $\fv[2] \gf \fv[2]$ are in $\phab$ notation, and
$r$ runs over the nine $\pph$ channels.

\section{Numerical results}
\label{sec:numerical-results}

\subsection{Numerical methods employed}
All numerical results presented in this section are computed for the AIM of the DMFT solution for a single-band, square
lattice Hubbard model with nearest neighbor hopping $\hopping = 1$, total density
$\density = 1.1$, inverse temperature $\invtemp = 10$, and local Coulomb interaction
$\interaction \in \{0.5, 1, 2, 3, 4\}$. Here, the DMFT solution is obtained with
w2dynamcis~\cite{w2dynamics}, a continuous-time quantum Monte Carlo (QMC) solver, which
also allows calculating the local two-particle Green's function and then, by employing the
parquet and Bethe--Salpeter equations, all local two-particle vertices. 
\footnote{Worm sampling \cite{Gunacker2015} and, for the one-particle Green's functions, symmetric improved estimators \cite{Kaufmann2019} are used.} The hybridization
expansion used by w2dynamcis is, however, not well suited in the limit of small
$\interaction$ since the hybridization becomes relatively large compared to the
interaction. This means that while the one-particle Green's function can still be obtained
with high enough accuracy and precision, the irreducible vertices are too noisy to be used
in the three-particle ladder.  Therefore, we employ different methods for computing $\iv$.

For the smallest interactions, $\interaction \in \{0.5, 1, 2\}$, we use a parquet equations
solver~\cite{KrienKauch2022,PhysRevResearch.3.013149} to compute the irreducible vertices
in the parquet approximation $\fiv \approx \interaction$, which yields noise free
results.
For $\interaction \in
\{3, 4\}$ we use the same parquet solver, but take the local fully irreducible vertex $\fiv$
from the QMC calculation as a starting point because it turns out that the resulting $\iv$
has less noise than the $\iv$ computed directly from the QMC results.

Further, exact diagonalization (ED) calculations are performed using the ED code of \cite{PhysRevResearch.4.033238}  with six bath sites. The latter are obtained from the
DMFT(QMC) solution via pole fitting~\cite{PhysRevB.103.045120}. A further alternative, which is however beyond the scope of the present
paper, would be to use the numerical renormalization group for
calculating the local two-particle vertex~\cite{Kugler2021a, Kugler2021b}.

In all cases the corresponding ladder diagrams are then evaluated for the AIM. That is, this way we can compute three-particle
local Green's functions and response functions of the AIM with the derived ladder approximation, and compare these against numerically calculated exact ones.

\subsection{Eigenvalues of \texorpdfstring{$M$'s}{M's}}

Before we compute any three-particle ladders we need to make sure that the geometric
series in \cref{eq:l} actually converges. To this end we compute the eigenvalues of the
$M$'s (i.e, $M$, $M_{12}$, $M_{23}$, and $M_{13}$) and check if their absolute values are
less than one. More precisely, when looking at \cref{eq:m12,eq:m23,eq:m13} (the
definitions of the $M$'s) we condense $(\nu, \omega, \sigma)$ and $(\nu', \omega',
\sigma')$ into two compound indices. For every $\nubar$ we then have a sparse square matrix ($\approx
\num{30000} \times \num{30000}$) for
which the eigenvalues are calculated. 

\begin{figure*}
    \centering
    \includegraphics[width=.7\textwidth]{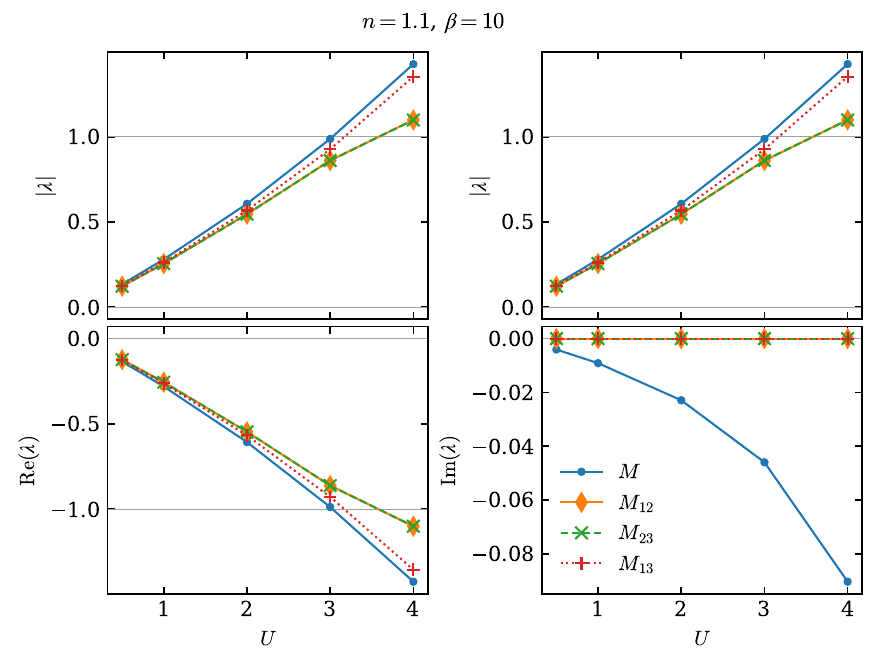}
    \caption{Largest absolute eigenvalues of $M$'s plotted over $\interaction$. }
    \label{fig:m-ev-over-u}
\end{figure*}

The absolute largest eigenvalues $\lambda$ over all values of $\nubar$ are shown in
\cref{fig:m-ev-over-u} for different interaction strengths $\interaction$.
We see that the eigenvalues of $M$ are larger than those of its components, and the
difference grows with $\interaction$. At $\interaction = 3$ the magnitude of the largest
eigenvalue of $M$ is just below one, so the ladder converges only up to about this point.
Despite that, we do not expect the ladder to blow up numerically because we compute the
ladder using the closed-form solution for the geometric series, $1/(1 - q)$, and the
eigenvalues are close to $-1$ and not $+1$. We also see that $M_{12}$ and $M_{23}$ yield
identical results which is not surprising since they both contain $\iv_{\ph}$ and are
mirror versions of each other. Furthermore, $M_{13}$ and therefore $\iv_{\pp}$ seem to
give the dominant contribution to the largest eigenvalue of $M$ and therefore the
three-particle ladder as a whole. The last thing to mention is that only $M$ seems to have
eigenvalues with an imaginary part. However, when looking at more than just the largest
eigenvalue we find that the components of $M$, especially $M_{13}$, have eigenvalues with
similar imaginary parts as well (not shown), so this is actually not an exceptional
result.

\subsubsection{\texorpdfstring{$\vrf[2]$}{chi2 vertex}: exact diagonalization vs
               approximate ladder}
\label{sssec:ed-vs-ladder} 
Next, we compare the ladder approximation
to the ED for three-particle vertex contributions to the second-order response functions.
Similarly as in \cite{chi2-paper}: First one-, two-, and three-particle correlators are
computed with the ED code~\cite{PhysRevResearch.4.033238}. Then the disconnected parts are
subtracted to obtain the full second-order response functions $\rf[2]$. Next the bubble
terms $\rf[2]_{0}$ and first-order terms $\rf[2]_{1}$ are computed from one- and
two-particle quantities (see \cref{fig:chi2-decomposition}).
Finally, $\rf[2]_{0}$ and $\rf[2]_{1}$ are subtracted from the full second-order response
function yielding the vertex term $\vrf[2]$. The difference to \cite{chi2-paper} is that
instead of the \enquote{physical} response functions $\rf_{\nnn}$, $\rf_{\nzz}$, and
$\rf_{xyz}$ we compute and compare the three spin components $\rf_{\uuu}$, $\rf_{\uud}$,
and $\rf_{\ububd}$. The reason for that is that these spin components are what we can
actually compute directly from the approximate ladder. As shown in \cite{chi2-paper} the
physical response functions are a linear combination of the spin components, and we do not
want that linear combination to hide or disguise potential differences between the ED
results and the ladder.

As explained at the end of \cref{sec:3p-ladder}, the contribution to $\vrf[2]$ from the
approximate ladder is computed in $\phab$ notation~\cref{eq:phia}. However, for better consistency and comparability
all results shown in this section use the same frequency notation as in \cite{chi2-paper},i.e., the $\ph$ notation from \cref{tab:g3-2p-frequency-notations}. That is, we have to translate back from \cref{eq:phia} to \cref{eq:omegaabc}.

\begin{figure*}
    \centering
    \includegraphics[width=17cm]{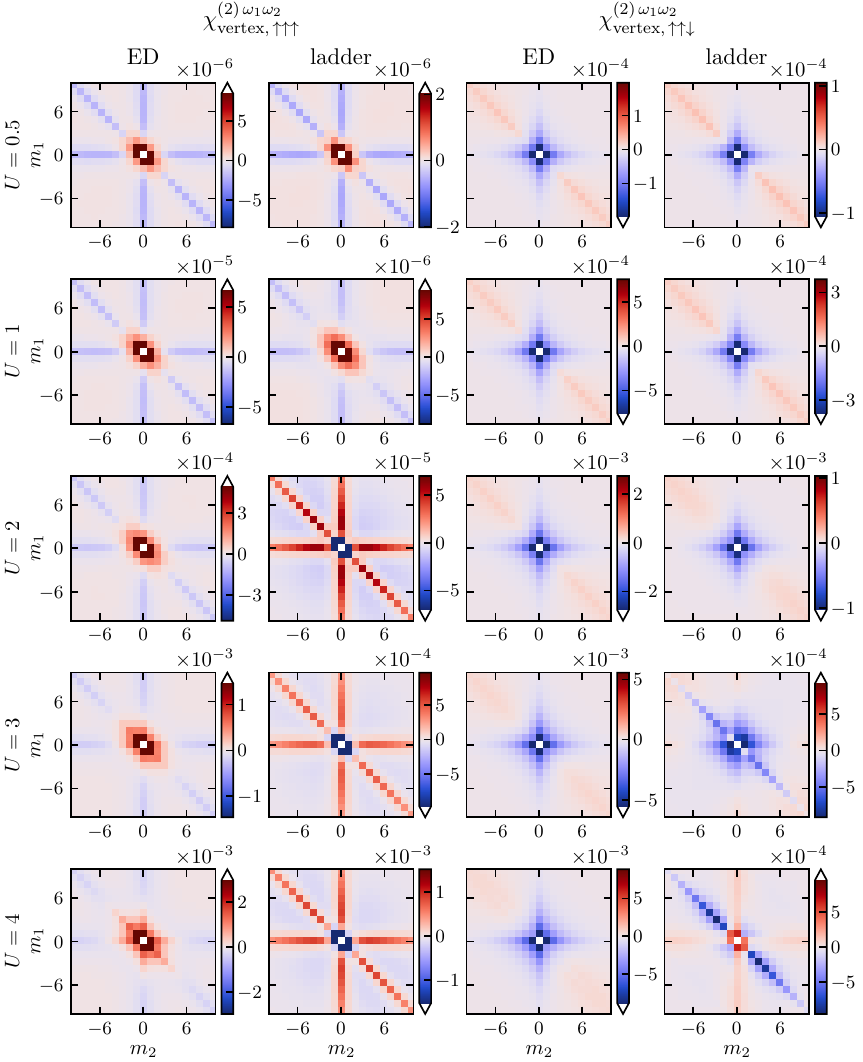}
    \caption{Comparison of spin component $\uuu$ and $\uud$ of $\vrf[2]^{\omega_1
             \omega_2}$ between ED and the approximate ladder for different values of the
             interaction $\interaction$ {(here and in the following figures inverse temperature is  $\invtemp =10$)}. The labels $m_i$ are the indices of the bosonic
             frequencies $\omega_i = 2 \pi m_i / \invtemp$. The static values ($\omega_1 =
             \omega_2 = 0$) are much larger than the other frequency components, so they
             are cut off from the color bars and instead shown in
             \cref{tab:ed-vs-ladder-1-4}.}
    \label{fig:ed-vs-ladder-1-4}
\end{figure*}

{\cref{fig:ed-vs-ladder-1-4}} compares the ED and ladder results of spin components $\uuu$
and $\uud$ of $\vrf[2]^{\omega_1 \omega_2}$ for different values of the on-site Coulomb
interaction $\interaction$.
We only plot the real parts since, as expected from the discussion at the end of
\cref{sec:3p-ladder}, the imaginary parts are at least seven orders of magnitude smaller
than the real part. The static values of the second-order
response functions, i.e. those at $\omega_1 = \omega_2 = 0$, are much larger (by
magnitude) than the other frequency components and would dominate the plot. Therefore,
they are cut off from the color bars and instead shown in \cref{tab:ed-vs-ladder-1-4}.
\begin{table}
    \centering
    \begin{tabular}{@{}lrrrr@{}}
        \toprule
        & \multicolumn{2}{c}{$\vrf[2]_{\uuu}^{0 0}$}
        & \multicolumn{2}{c}{$\vrf[2]_{\uud}^{0 0}$}
        \\
        \cmidrule(lr){2-3} \cmidrule(l){4-5}
        U & \multicolumn{1}{c}{ED} & \multicolumn{1}{c}{ladder}
          & \multicolumn{1}{c}{ED} & \multicolumn{1}{c}{ladder}
        \\
        \midrule
        \num{0.5} & \num{ 5.6e-05} & \num{ 1.5e-05} & \num{-6.3e-04} & \num{-3.4e-04}
        \\
        \num{1.0} & \num{ 4.0e-04} & \num{ 4.8e-05} & \num{-2.3e-03} & \num{-1.1e-03}
        \\
        \num{2.0} & \num{ 2.4e-03} & \num{-9.1e-04} & \num{-7.1e-03} & \num{-2.2e-03}
        \\
        \num{3.0} & \num{ 5.5e-03} & \num{-8.9e-03} & \num{-1.2e-02} & \num{ 9.5e-04}
        \\
        \num{4.0} & \num{ 7.5e-03} & \num{-1.5e-02} & \num{-1.6e-02} & \num{ 5.7e-03}
        \\
        \bottomrule
    \end{tabular}
    \caption{Static values ($\omega_1 = \omega_2 = 0$) of $\vrf[2]$ cut off from the color
             bars of \cref{fig:ed-vs-ladder-1-4}}
    \label{tab:ed-vs-ladder-1-4}
\end{table}

The first thing we see when looking at \cref{fig:ed-vs-ladder-1-4} is that the approximate
ladder yields qualitatively good results only for small values of the interaction. For the
$\uuu$ component this is until $\interaction = 1$; for the $\uud$ component it is until
$\interaction = 3$. Even then, however, the results are off by a factor of two to ten, so
quantitatively the approximate ladder is not good.

\begin{figure*}
    \centering
    \includegraphics[width=17cm]{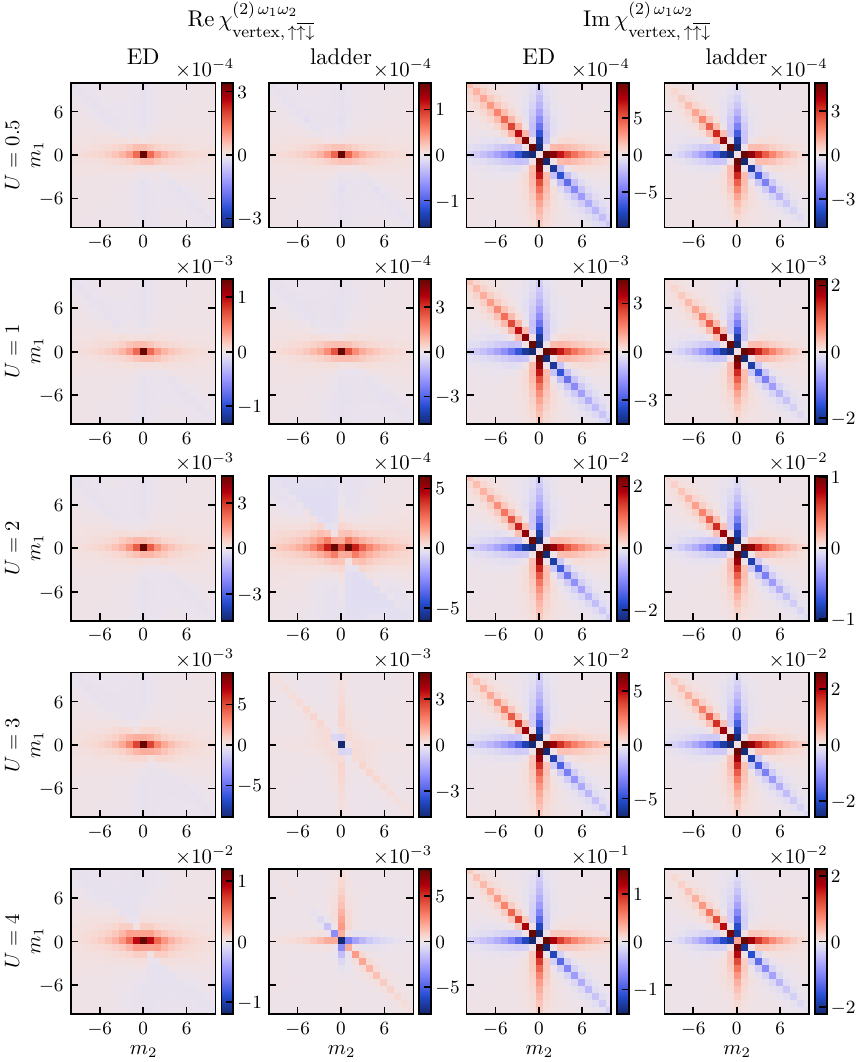}
    \caption{Comparison of the real and imaginary parts of spin component $\ububd*$ of
             $\vrf[2]^{\omega_1 \omega_2}$ between ED and the approximate ladder for
             different values of the on-site Coulomb interaction $\interaction$. The axis
             labels $m_i$ are the indices of the bosonic frequencies $\omega_i = 2 \pi m_i
             / \invtemp$.}
    \label{fig:ed-vs-ladder-7}
\end{figure*}

\Cref{fig:ed-vs-ladder-7} compares the real and imaginary parts of the ED and ladder
results of spin component $\ububd*$ of $\vrf[2]^{\omega_1 \omega_2}$ for different values
of the interaction $\interaction$.
Similar to the spin component $\uuu$ the approximate ladder only yields qualitatively good
results until $\interaction = 1$ -- at least for the real parts. The imaginary parts look
fine for all values of $\interaction$. However, looking at the color bars reveals that
quantitatively the ladder results for spin component $\ububd$ are  just as bad as those for
the others.

The fact that for all spin components, the qualitatively good-looking results are
consistently too small is no surprise. After all, we average over all nine $\pph$ channels
instead of summing. Of course, as we have shown at the end of \cref{sec:3p-ladder},
summing would overcount diagrams with few vertices, but we conjecture that ladder terms of
high enough order, i.e., number of vertices, are exclusive to a channel. Averaging
therefore cuts down higher order contributions by a factor of nine.

\subsection{Approximate ladder in different channels}

\begin{figure*}
    \centering
    \includegraphics[width=14cm]{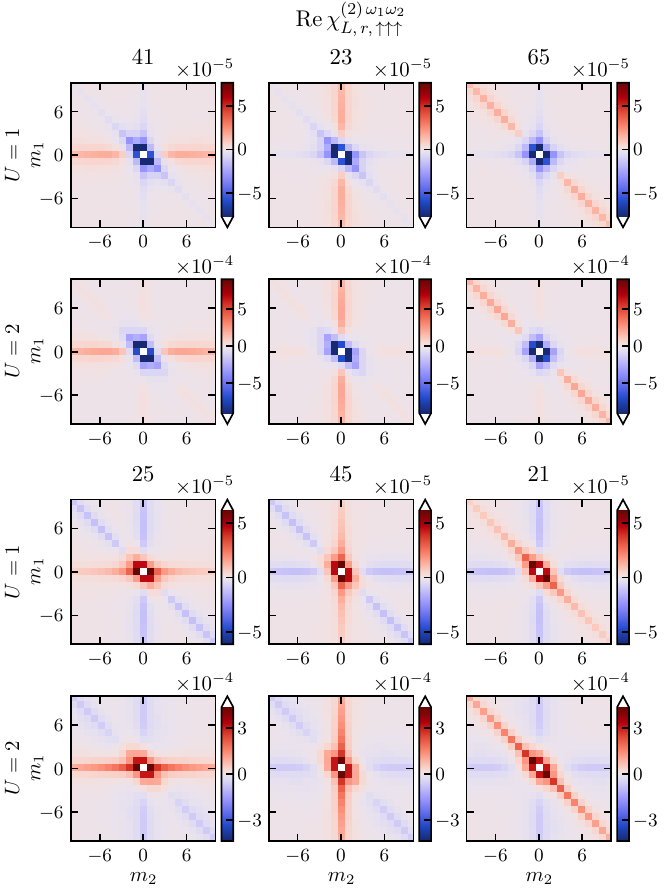}
    \caption{Comparison of the contributions of the approximate ladder $L$ in six $\pph$
             channels $r \in \{41, 23, 65, 25, 45, 21\}$ to the $\uuu$ component of the
             second-order response function $\rf[2]_{L, r}^{\omega_1 \omega_2}$ for
             Coulomb interactions $\interaction = \{1, 2\}$. The labels $m_i$ are the
             indices of the bosonic frequencies $\omega_i = 2 \pi m_i / \invtemp$. As in
             \cref{fig:ed-vs-ladder-1-4}, the static components ($\omega_1 = \omega_2 =
             0$) are cut off and shown in \cref{tab:chi2-ladder-channels} instead.}
    \label{fig:chi2-ladder-channels}
\end{figure*}

In this section we take a closer look at the contributions of the different $\pph$
channels $r$ to the second-order response function computed only from the approximate
ladder $L$, i.e.,
\begin{equation}
    \rf_{L, r}^{\omega_1 \omega_2}
        = \sum_{\nu_1 \nu_2 \nu_3} L_{r}^{\nu_1 \omega_1 \nu_2 \omega_2 \nu_3}.
\end{equation}
We already pointed out at the end of \cref{sec:3p-ladder} that six of the nine channels
form mirror pairs. These pairs have the same real part but opposing imaginary parts, which
all cancel eventually. Therefore, only six channels are of interest.
\Cref{fig:chi2-ladder-channels} compares the spin component $\uuu$ of the real parts of
$\rf_{L, r}^{\omega_1 \omega_2}$ in six such channels at interaction strengths
$\interaction = 1$ and $\interaction = 2$.
As in \cref{fig:ed-vs-ladder-1-4} the static frequency components ($\omega_1 = \omega_2 =
0$) would dominate the plot and are therefore cut off and presented separately in
\cref{tab:chi2-ladder-channels}.
\begin{table*}
    \centering
    \begin{tabular}{@{}lrrrrrr@{}}
        \toprule
            & \multicolumn{6}{c}{$\rf_{L, r, \uuu}^{0 0}$}
        \\
        \cmidrule{2-7}
        $U$ & \multicolumn{1}{c}{$r = 41$} & \multicolumn{1}{c}{$r = 23$}
            & \multicolumn{1}{c}{$r = 65$} & \multicolumn{1}{c}{$r = 25$}
            & \multicolumn{1}{c}{$r = 45$} & \multicolumn{1}{c}{$r = 21$}
        \\
        \midrule
        \num{1.0} & \num{-3.3e-04} & \num{-3.3e-04} & \num{-3.3e-04}
                  & \num{ 2.6e-04} & \num{ 2.6e-04} & \num{ 2.6e-04}
        \\
        \num{2.0} & \num{-4.4e-03} & \num{-4.4e-03} & \num{-4.4e-03}
                  & \num{ 1.2e-03} & \num{ 1.2e-03} & \num{ 1.2e-03}
        \\
        \bottomrule
    \end{tabular}
    \caption{Static values ($\omega_1 = \omega_2 = 0$) of $\rf[2]_{L, \, r, \, \uuu}$ cut
             off from the color bars of \cref{fig:chi2-ladder-channels}}
    \label{tab:chi2-ladder-channels}
\end{table*}

When looking at \cref{fig:chi2-ladder-channels} we see that all results share two
features: First, the main contributions are around the center; second, one of the three
lines $\omega_i = 0$, $i \in \{1, 2, 3\}$, is more pronounced than the others. The latter
is not surprising, since the Feynman diagrams for a single ladder have a distinguished
direction. This also highlights once more why averaging over the approximate ladders in
all channels is important if we want the get the right structure of the three-particle
vertex.

We also see that the contributions from the upper and lower three channels in
\cref{fig:chi2-ladder-channels} have opposite signs. For $\interaction = 1$ they even have
a very similar magnitude. However, since for each of the lower three channels there is a
second one with the same real part, those contributions outweigh the upper three channels.
The final result for $\vrf[2]_{\uuu}$ at $\interaction = 1$ shown in
\cref{fig:ed-vs-ladder-1-4} has therefore positive values around the center.

When comparing the results between $\interaction = 1$ and $\interaction = 2$ in
\cref{fig:chi2-ladder-channels} we notice that while the plots still look qualitatively
similar, the balance in magnitude shifts. The static frequency components in
\cref{tab:chi2-ladder-channels} show this best. While the values for the first three
channels rise by a factor of more than ten, those of the last three channels only rise by
a factor of less than five. In absolute terms, the channels $41$, $23$, and $65$ now
outweigh the other channels by a factor of more than three. Therefore, the sign of the
total result for $\vrf[2]_{\uuu}$ shown in \cref{fig:ed-vs-ladder-1-4} changes at
$\interaction = 2$.

%% file: conclusion_and_outlook.tex
\section{Conclusion and outlook}
\label{sec:conclusion-and-outlook}

We have generalized the Bethe--Salpeter equations from two to three particle ladders.
Due to the increased complexity of three-particle diagrams when it comes to the property
of reducibility, the resulting equations are much more involved than on the two-particle
level. Also, the fully irreducible building blocks of the exact three-particle ladder are more complicated,
and we do not yet have a feasible way of computing them. For this reason we have derived
an approximate ladder, built simply from irreducible two-particle vertices $\iv$ and full
Green's functions. Admittedly, even this approximation turned out to be quite complex and involved, 
but  yields a viable way for approximately computing the
full three-particle vertex $\fv[3]$. Our three-particle ladder equations bear some similarities to the Faddeev equations~\cite{Faddeev1961} for the three-body problem; they are however somewhat more involved since we do not keep the particle(hole) number fixed in-between the three incoming and the three outgoing particles(holes).

If we knew the 3PI vertex in a given channel exactly\footnote{More precisely, we would also need the two-particle vertex for the 1PR diagrams.},  the corresponding ladder equation in that channel 
is exact and thus yields the exact full three-particle vertex. But, we do not
have a feasible way of computing that 3PI  vertex yet. An approximation will most likely be
necessary. In the spirit of diagrammatic extensions of DMFT~\cite{RMPVertex}, it might be easier to find a good approximation or a numerical way to calculate 
the {\em local}  irreducible
vertex than for the full one. Here, we employ an approximation
where the 3PI vertex is built up from the three irreducible two-particle vertices that each connect two of the particle(hole) lines.

The numerical computations show that the thus approximated ladder only yields  good
results for very small values of the local Coulomb interaction $\interaction$ and even then only qualitatively. At least
this holds for the investigated parameters and for the non-linear response function studied. 
There might be parameter regimes where the approximation
is generally better; or quantities where we do not combine fermionic lines from both ends of the ladder.
For example, two-particle ph ladders provide a quite good approximation for the magnetic susceptibility,
if expressed in terms of $\iv$ instead of the bare Coulomb interaction. But there, we combine the particle and hole line from one side of the ladder to a bosonic (para)magnon line. Calculating instead the $\phb$ ladder and then combining particle and hole lines from opposite sides of the ladder would be a rather bad approximation for the magnetic susceptibility.
Similarly for a three-particle quantity where we do not have to average over ladders in all nine channels
because the physics happens predominantly in one \enquote{direction} might also give better results for the three-particle ladder. One could also look at a better way
to avoid the double counting issues that arise from summing the nine channels. 
{This would require an additional major effort beyond the scope of the present paper. That is, one would need to identify all diagrams that are included twice or more in different three-particle ladders (including at least all second order diagrams) and subtract them from the added contributions of the nine channels.}

In the end, computing three-particle vertices is still in its very early stages of
development. So while the current results fall short of expectations, they are still
useful and a good first step that can be built upon.

\begin{acknowledgments}
    We thank M.~Pickem and M.~Wallerberger  for valuable and helpful discussions. 
 We acknowledge funding through the Austrian Science Fund (FWF) projects  P32044  
 and V1018.
 Calculations have been done in part on 
 the Vienna Scientific Cluster (VSC). 

For the purpose of open access, the authors have applied a CC BY public copyright license to any author accepted Manuscript version arising from this submission.

The code for the  calculation of the three-particle ladder is available under open source license at 
\url{https://gitlab.tuwien.ac.at/e138/e138-01/software/braids}.

\end{acknowledgments}

%% file: acknowledgments.tex

%% file: supplemental.tex

\newcommand{\qt}{\textquotedbl}
\newcommand{\IPR}{\textcolor{red}{1PR}}
\newcommand{\IIPR}{\textcolor{blue}{2PR}}
\newcommand{\IIIPR}{\textcolor{magenta}{3PR}}
\pagebreak
\clearpage
\onecolumngrid
\begin{center}
  \textbf{\large Supplemental Material:  Ladder equation for the three-particle vertex and its approximate solution}\\[.2cm]

{Patrick Kappl, Tin Ribic, Anna Kauch, Karsten Held}\\[.1cm]

{\it Institute of Solid State Physics, TU Wien, 1040 Vienna, Austria}\\

(Dated: \today)\\[.5cm]

\begin{minipage}{15cm}
This Supplemental Material provides a closer look at the (ir)reducibility of the three-particle vertex. Diagrams 
for the three-particle vertex can be one-, two-, and three-particle reducible  as well as fully three-particle irreducible.
Things become quite involved since diagrams can be reducible in  more than one way.
Nonetheless, by properly subtracting all other diagrams, we can --in principle-- determine the  fully three-particle irreducible three-particle 
vertex or the one that is three-particle reducible in a given channel.
\\[.5cm]
\end{minipage}
\end{center}


\setcounter{equation}{0}
\setcounter{figure}{0}
\setcounter{table}{0}
\setcounter{section}{0}
\setcounter{page}{1}
\renewcommand{\theequation}{S\arabic{equation}}
\renewcommand{\thefigure}{S\arabic{figure}}
\renewcommand{\bibnumfmt}[1]{[S#1]}
\renewcommand{\thesection}{S\Roman{section}}

\begin{figure}[H]
\centering
\includegraphics[width=0.5\linewidth]{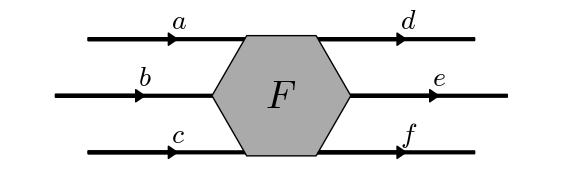}
\caption{\label{3PFV} Notational convention for the three-particle vertex within the SM. The external propagators are depicted for reasons of clarity only, they are not part of the vertex. The full vertex $F$ contains all Feynman diagrams which connect all incoming and outgoing particle-lines.}
\end{figure}

While the properties of the two-particle vertex are already quite well established and have been investigated closely \cite{Rohringer2012,RMPVertex}, the three-particle vertex is less well understood and the topic of the main manuscript.
In this Supplemental Material (SM), we provide a more detailed overview of the features of three particle vertices that augments the discussion of the main text.  This SM is self-contained, notations differ sometimes from that of the main text. We will restrict ourselves to systems with particle-number conservation. We will start with a systematic classification of three-particle vertices in terms of their (ir)reducibility. The notation will be chosen in agreement with figure \ref{3PFV}. The indices $a$, $b$, ...  are understood to be multi-indices carrying all necessary information about the involved incoming and outgoing states (in the main text we used labels 1, 2, ...).
\section{One-particle reducible contributions}
\label{3part1PR}
\begin{figure}[H]
\centering
\includegraphics[width=0.5\linewidth]{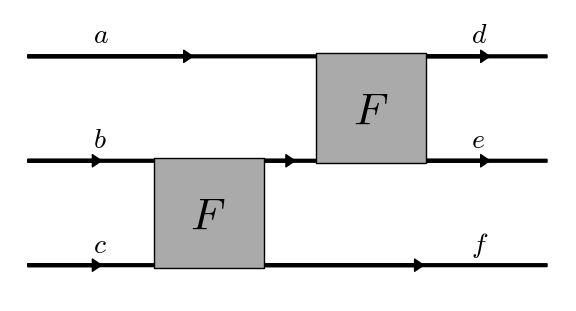}
\caption{\label{3P1PRI} Here, we show all diagrams contributing to the three-particle vertex that are \IPR \ in the channel separating $a , d ,e$ from $b , c , f$. It can be written as a combination of two two-particle vertices and a Green's function line connecting them. The outer legs are shown for clarity.}
\end{figure}

Unlike the two-particle vertex, the three-particle vertex is not precluded from containing one-particle reducible (\IPR) contributions by conservation of particle number alone. Here, a diagram is said to be \IPR \ if it can be disconnected into two two-particle diagrams by cutting a single one-particle propagator. The structure of \IPR \ contributions to the full vertex is very simple if one decides to work with dressed one-particle propagators already containing the self-energy. There is only one type of diagram which contributes to the three-particle vertex and is \IPR, see figure \ref{3P1PRI}.
Since two-particle vertices are always one-particle irreducible, no overcounting problems arise for the diagrams shown in figure \ref{3P1PRI}. There are $9$ different channels in which a three-particle vertex can be \IPR \ as given in table \ref{1PRchanneltable}.
\begin{table}[H]
 \centering
  \begin{tabular}{ | l | c | r |}
    \hline
    $a,b ,d$ & $a,b ,e$ & $a,b ,f$\\
    $c,e ,f$ & $c,d ,f$ & $c,d ,e$\\ \hline
    $a,c ,d$ & $a,c ,e$ & $a,c ,f$\\
    $b,e ,f$ & $b,d ,f$ & $b,d ,e$\\ \hline
    $b,c ,d$ & $b,c ,e$ & $b,c ,f$\\
    $a,e ,f$ & $a,d ,f$ & $a,d ,e$\\ \hline
  \end{tabular}
\caption{\label{1PRchanneltable}Systematic listing of all channels of one-particle reducibility for three-particle vertices. Each entry has two lines, the outer legs in every line remain connected to each other when the one-particle propagator is cut, while the connection between the legs of the upper and lower line is broken.}
\end{table}
One might be surprised by the number of channels, as there are $10$ different ways of grouping $6$ elements into two sets of $3$ elements. However, one of these $10$ decompositions cannot be conducted. The missing decomposition is $a,b,c$ and $d,e,f$ {(corresponding to the $ppp$ channel)}. This sort of diagram cannot exist due to particle number conservation; corresponding diagrams would have to be separable into two-particle vertices with either $3$ entering lines and $1$ leaving line or $4$ entering and no leaving lines and vice-versa---an impossibility due to conservation of particle number. We further take note of the fact that a three-particle diagram can be reducible in one one-particle channel at most. If this were not the case, after the first cut, we would end up with two two-particle diagrams.
{Now, the second cut of one line would necessarily cut one of these  two-particle diagrams while the other two-particle diagram would need to  decompose even without a cut being performed, i.e. be disconnected, in order to  separate the distinct legs of the second \IPR \ channel.} Two-particle vertices are however always one-particle irreducible\footnote{For systems with particle-number conservation}. This proves by contradiction that a diagram can only be \IPR \ in one channel. This also means that no double-counting corrections are necessary when eliminating  \IPR \ contributions in the nine channels of Tabel~\ref{1PRchanneltable} from the full three-particle-vertex. 
\section{Two-particle reducible contributions}
\begin{figure}
\centering
\includegraphics[width=0.55\linewidth]{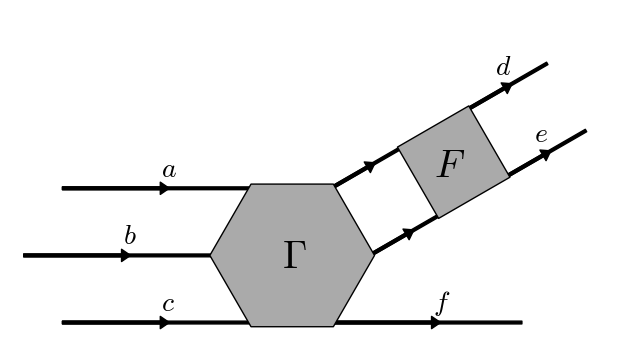}
\caption{\label{3P2PR} All diagrams contributing to the three-particle vertex and \IIPR \ in the channel separating $d ,e$ from $a,b,c,f$}
\end{figure}

\begin{figure}[H]
\centering
\includegraphics[width=0.45\linewidth]{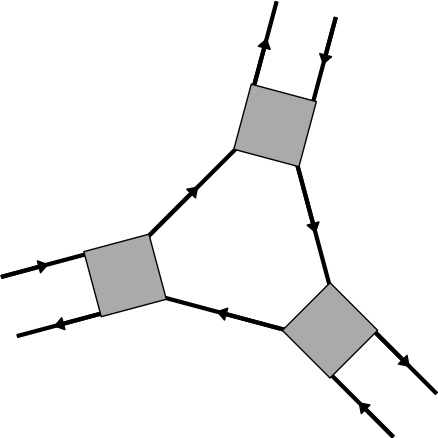}
\caption{\label{coollogo} Diagrams of this structure are \IIPR \ in three $ph$-like channels, but not necessarily simultaneously reducible in more than one.}
\end{figure}

\begin{figure}[H]
\centering
\includegraphics[width=0.65\linewidth]{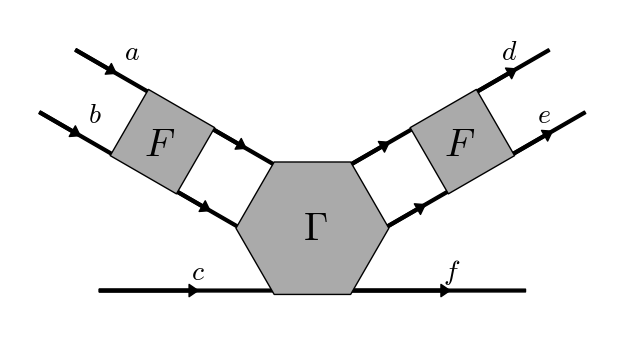}
\caption{\label{3P22PR} All diagrams contributing to the three-particle vertex and simultaneously \IIPR \ in the channels separating $a,b$ and $d ,e$}
\end{figure}

There is a plethora of two-particle reducible (\IIPR) diagrams included in the full three-particle vertex. A three-particle diagram is considered \IIPR \ if it can be disconnected into a three-particle and a two-particle diagram by cutting two internal one-particle propagators. The possibility of extracting an internal propagator with a self-energy insertion, thereby generating a one-particle diagram and a four-particle diagram is explicitely excluded. The number of channels increases dramatically compared to the two-particle level. In principle, there are $15$ simple ways to be two-particle disconnected on the three-particle level, $6$ $pp$-like ones and $9$ $ph$-like ones. They are defined by which two outer legs can be disconnected from the remaining four by cutting two internal propagators. A table of all channels of \IIPR \ diagrams is given below. We will refer to channels by the pair of outer legs which is separated from the rest.
\begin{table}[H]
 \centering
  \begin{tabular}{ | c | c | c | c | c |}
    \hline
    $a,b$     & $d,e$     & $a,d$     & $b,d$     & $c,d$     \\
    $c,d,e,f$ & $a,b,c,f$ & $b,c,e,f$ & $a,c,e,f$ & $a,b,e,f$ \\ \hline
    $a,c $    & $d,f$     & $a,e$     & $b,e$     & $c,e$     \\
    $b,d,e,f$ & $a,b,c,e$ & $b,c,d,f$ & $a,c,d,f$ & $a,b,d,f$ \\ \hline
    $b,c$     & $e,f$     & $a,f$     & $b,f$     & $c,f$     \\
    $a,d,e,f$ & $a,b,c,d$ & $b,c,d,e$ & $a,c,d,e$ & $a,b,d,e$ \\ \hline
  \end{tabular}
\caption{\label{2PRchanneltable}Systematic listing of all channels of two-particle reducibility for three-particle vertices. Each entry has two lines, the outer legs in every line remain connected to each other when two one-particle propagators are cut.}    
\end{table}The structure of diagrams reducible in the $pp$-like channel separating $d,e$ and $a,b,c,f$ is given in figure \ref{3P2PR}. 
The major problem in attempting a systematic classification of \IIPR \ three-particle diagrams comes from the diversity of combinations of reducibility. Just because a diagram is \IIPR \ in a given channel does not necessarily require it to be irreducible in the other ones, as is the case for the two-particle vertex. There are some restricting features, however. A good example for the arising difficulties are all the diagrams of the structure depicted in figure \ref{coollogo}. These diagrams additionally feature non-simultaneous reducibility, which is discussed in more detail in section \ref{RSOGK}.
To further our understanding of the structure of two-particle reducible contributions to the three-particle vertex, we investigate restrictions on reducibility. To this end, the standard argument against multiple reducibility is applied: Let us consider a diagram which is reducible in two different channels, $i,j$ and $k,l$ with
\begin{equation*}
 i,j,k,l \in  \{ a,b,c,d,e,f \} \; \vert \; i \neq j , k \neq l , \{ i,j \} \neq \{ k,l \}. 
\end{equation*}
This means that there are two sets of cuts disconnecting the diagram in specific manners. We imagine applying both sets of cuts. The first set disconnects the diagram into a two-particle diagram with two ($i,j$) of the original six outer legs and a three-particle diagram with the remaining four. The second set of cuts should now disconnect two more outer legs ($k,l$) from the rest. A problem might arise if both sets of cuts contain cutting the same line as it would be the case with some diagrams of the structure depicted in figure \ref{coollogo}, but this case will be treated separately in section \ref{RSOGK}. Excluding the 'shared cut' case, no complications occur if $i,j$ and $k,l$ do not share an element, i.e.,
$\nexists \, m \; \vert \; m \in \{ i,j \} , \, m \in \{ k,l \}$.
and the diagrams are of the generalized structure depicted in figure \ref{3P22PR}. Note that the opposite case, i.e.,
$\exists \, m \; \vert \; m \in \{ i,j \} , \, m \in \{ k,l \}$
leads to a contradiction, as will be shown in section \ref{CompIssue}. An analogous line of reasoning can be followed for the case of threefold two-particle reducibility, yielding similar results.
\subsection{Incompatible channels}
\label{CompIssue}
There are some restrictions on in which channels a diagram can be reducible at the same time.
Let us consider pairs of \IIPR \ channels which share an external line index, for example $a,b$ and $b,c$. After the first set of cuts (no matter which one), we have disconnected our diagram into a two-particle and a three-particle part. If the second set of cuts was to be applied applied as well, one of the outer legs of the three-particle contribution as well as of the two-particle contribution would have to be disconnected from their respective remaining diagrams. This can work neither for the two-particle nor for the three-particle part. The implication is that a given diagram cannot be reducible in two channels which share an index, unless performing the cuts to disconnect the diagram in the channels is impossible for some reason, as will be discussed \ref{CompIssue}. We decide to call such channels incompatible with respect to two-particle reducibility.

Each channel is incompatible with $8$ and compatible with $6$ other channels. In total, there are $45$ distinct pairs of compatible channels and $15$ distinct sets of three compatible channels. There are $4$ distinct types of pairs of compatible channels (incoming and outgoing $pp$-like, incoming $pp$ and $ph$, outgoing $pp$ and $ph$, as well as double $ph$ with multiplicities $9,9,9$ and $18$ respectively). There are $2$ types of triplets of reducibilities, $pp$-$pp$-$ph$ and $ph$-$ph$-$ph$, appearing with multiplicities $9$ and $6$. Every two-particle channel appears in exactly three triplets.

\begin{figure}[H]
\centering
\includegraphics[width=0.65\linewidth]{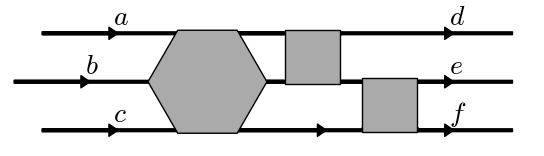}
\caption{\label{2PMasking} Reducibility in the $e,f$ channel masking reducibility in the $d ,e$ channel.}
\end{figure}

Incompatible two-particle channels display additional behaviour hampering our attempts at finding fully irreducible three-particle vertices: reducibility on the two-particle level can be masked. Consider a diagram of the structure depicted in figure \ref{2PMasking}. Obviously, the diagram is reducible in $e,f$, but would become reducible in $d,e$ upon removing the $e,f$ \IIPR \ contribution. Such diagrams with masked two-particle reducibility are always \IIIPR \ as well, so taking some care in removing them is required to avoid overcounting issues.\footnote{This masking phenomenon already appears on the two-particle vertex level, which is the basis of the parquet formalism.~\cite{Bickers2004}}
\subsection{Shared cuts}
\label{RSOGK}
\begin{figure}[H]
\centering
\includegraphics[width=0.65\linewidth]{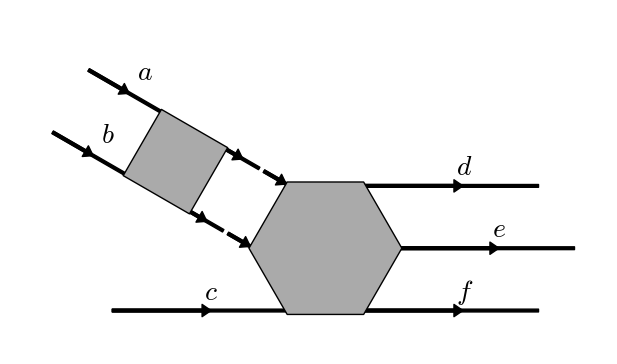}
\caption{\label{2PdisconnNonSim} Remaining structure of \qt shared cut\qt \ diagram after disconnecting $a,b$}
\end{figure}

One can easily see that while the diagrams in figure \ref{coollogo} are \IIPR \ in $3$ different channels, they are not of the structure shown in figure \ref{3P22PR}. The issue of diagrams not simultaneously separable needs to be resolved. We start from a diagram which is \IIPR \ in two compatible channels, but not simultaneously reducible in both. This can only happen if the first set of cuts somehow interferes with the seccond cut if both cuts were to be applied. The only possibility of such an event occurring is when the same propagator would be cut by both sets of cuts. Without loss of generality, let us assume that the compatible channels in question are $a,b$ and $d,e$. We first conduct the cuts required to separate $a$ and $b$ from the remainder and end up with something akin to figure \ref{2PdisconnNonSim}.
Note that the directions of the one-particle propagators in figure \ref{2PdisconnNonSim} do not matter for our argument. If the cuts required to disconnect $d,e$ could be performed on the three-particle part, the diagram would be simultaneously separable in both channels and thus contained in the ones depicted in figure \ref{3P22PR}. This means that at least one cut which is necessary to disconnect $d,e$ is in some way interfered with by the first set of cuts. Let us first consider the (theoretical) possibility that one (or both) of the $d,e$ cuts are located within the two-particle part of the disconnected diagrams. A single cut within the two-particle part is not sufficient to disconnect any line from any other. A single cut within the three-particle part is not sufficient to disconnect two external legs from the remaining ones, though it is possible to disconnect the three-particle part into two two-particle pieces. Disconnecting $d,e$ from $c,f$ in the three-particle part, with each pair remaining connected to one of the legs originating from the $a,b$ cut is the only way of achieving a separation of $d,e$ from $c,f$. The problem with this sort of decomposition is, that such a diagram would not be \IIPR \ in $d,e$ in the first place, because performing only the $d,e$ cuts would leave the original diagram fully connected (because the two-particle propagator with $a,b$ cannot be disconnected by a single cut, acting as a chain-link for the diagram). The last remaining possibility is a \qt shared cut\qt -a single particle line which is included in both sets of cuts. In this case only a single line is cut when disconnecting $d,e$ after $a,b$, the remaining three-particle part after the $a,b$ cuts being \IPR \ in a channel which allows disconnecting $d,e$ from $c,f$. This implies that all diagrams which are \IIPR \ but not simultaneously \IIPR \ in two channels are actually of the structure in figure \ref{coollogo} and actually \IIPR \ --but not simultaneously \IIPR-- in exactly three compatible channels.
\section{Three-particle reducible contributions}
\begin{figure}[H]
\centering
\includegraphics[width=0.85\linewidth]{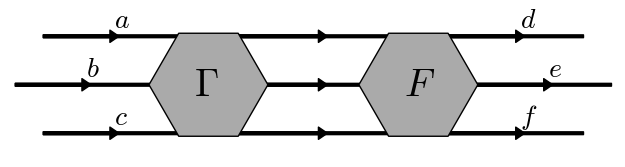}
\caption{\label{3P3PR} Structure of all diagrams contributing to the three-particle vertex and \IIIPR \ in the channel separating $a,b,c$ from $d,e,f$}
\end{figure}

\begin{figure}
\centering
\includegraphics[width=0.85\linewidth]{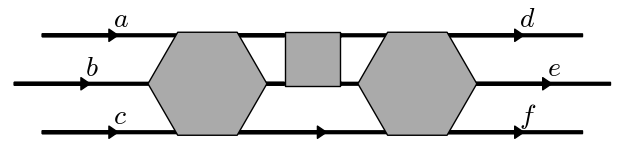}
\caption{\label{3P3PRProblem} Schematic representation of the issues arising from the interplay between two- and three-particle reducibility.}
\end{figure}

Even more complicated than the case of \IPR \ and \IIPR \ diagrams, three-particle reducibility (\IIIPR) offers a wide range of different channels and interplay with lower order reducibility. The channels can be labeled in accordance with the \IPR \ case, with one additional channel being introduced, the one disconnecting $a,b,c$ from $d,e,f$, which is not prohibited due to particle number conservation any longer when cutting three single-particle propagators. The general structure of the reducible diagrams is given in figure \ref{3P3PR}.
We also note that only diagrams which can be disconnected into two three-particle pieces are considered to be \IIIPR \ here. There is a possibility of disconnecting a three-particle diagram into a two-particle and a four-particle part by cutting three lines. This sort of decomposition is ignored for the sake of reducibility because it would relate the three-particle vertex to the four-particle one. All three-particle diagrams are reducible in this manner for a system with only quartic interaction. This is consistent with the definition of two-particle (ir)reducibility.
\\Note that the notion of an irreducible vertex becomes blurred when advancing from the two-particle to the three-particle level. At first glance, one might assume that $\Gamma$, as depicted in figure \ref{3P3PR} contains all diagrams irreducible in the $a,b,c$-$d,e,f$ channel. Considering a diagram of the structure depicted in figure \ref{3P3PRProblem}, it is clear that the two-particle insert has to be classified as a part of the full three-particle vertex. If such \IIPR \ contributions are considered within $\Gamma$ an overcounting problem is the consequence. Therefore, in constructing a Bethe-Salpeter-like equation for the three-particle level, these \IIPR \ contributions need to be explicitly taken into account.
\\An immediate consequence of the \IIPR \ contributions between the 'proper' three-particle vertices is that the two possible ways of constructing Bethe-Salpeter like equations are not fully equivalent; unlike at the two-particle level, the order of $\Gamma$ and $F$ matters. Also, the irreducible vertex $\Gamma$ depicted in figure \ref{3P3PR} is not only irreducible in the three-particle particle $a,b,c-d,e,f$ channel, but also in the two-particle $d,e$, $e,f$ and $d,f$ channels. Exchanging the order of $\Gamma$ and $F$ within the Bethe-Salpeter like equation would require another $\Gamma$ irreducible in the three-particle particle $a,b,c-d,e,f$ channel and the two-particle $a,b$, $b,c$ and $a,c$ channels. Therefore, the Bethe-Salpeter like equations \footnote{We chose one of two possible orders of $F$ and $\Gamma$, but there is another, equivalent, way of writing the equations.} on the three-particle level are of the structure
\begin{multline}
F \begin{pmatrix} a & b & c \\ d & e & f \end{pmatrix} = \Gamma \begin{pmatrix} a & b & c \\ d & e & f \end{pmatrix} 
 + \dfrac{1}{2} \sum_{1,2} \Gamma \begin{pmatrix} a & b & c \\ 1 & 2 & f \end{pmatrix} G(1) G(2) \; F \begin{pmatrix} 1 & 2 \\ d & e \end{pmatrix} \\
 + \dfrac{1}{2} \sum_{1,2} \Gamma \begin{pmatrix} a & b & c \\ d & 1 & 2 \end{pmatrix}  G(1) G(2) \; F \begin{pmatrix} 1 & 2 \\ e & f \end{pmatrix} 
 + \dfrac{1}{2} \sum_{1,2} \Gamma \begin{pmatrix} a & b & c \\ 1 & e & 2 \end{pmatrix}  G(1) G(2) \; F \begin{pmatrix} 1 & 2 \\ d & f \end{pmatrix} \\
 + \dfrac{1}{6} \sum_{1,2,3} \Gamma \begin{pmatrix} a & b & c \\ 1 & 2 & 3 \end{pmatrix}  G(1) G(2) G(3) \; F \begin{pmatrix} 1 & 2 & 3 \\ d & e & f \end{pmatrix} .
\label{BS3pppMain}
\end{multline}
for the $ppp$-like channel and
\begin{multline}
F' \begin{pmatrix} a & b & c \\ d & e & f \end{pmatrix} = \Gamma \begin{pmatrix} a & b & c \\ d & e & f \end{pmatrix} 
 + \dfrac{1}{2} \sum_{1,2} \Gamma \begin{pmatrix} a & b & c \\ 1 & 2 & f \end{pmatrix} G(1) G(2) \; F \begin{pmatrix} 1 & 2 \\ d & e \end{pmatrix} \\
 + \dfrac{1}{1} \sum_{1,2} \Gamma \begin{pmatrix} a & b & 2 \\ d & 1 & f \end{pmatrix}  G(1) G(2) \; F \begin{pmatrix} 1 & c \\ e & 2 \end{pmatrix} 
 + \dfrac{1}{1} \sum_{1,2} \Gamma \begin{pmatrix} a & b & 2 \\ 1 & e & f \end{pmatrix}  G(1) G(2) \; F \begin{pmatrix} 1 & c \\ d & 2 \end{pmatrix} \\
 + \dfrac{1}{2} \sum_{1,2,3} \Gamma \begin{pmatrix} a & b & 3 \\ 1 & 2 & f \end{pmatrix}  G(1) G(2) G(3) \; F' \begin{pmatrix} 1 & 2 & c \\ d & e & 3 \end{pmatrix} .
\label{BS3pphMain}
\end{multline}
for $pph$-like channels (The example is given for the $a,b,f-d,e,c$ channel specifically.). For above equations, the variables of the vertices were written as a matrix. The upper line of variables denotes incoming frequencies, the lower line denotes outgoing ones. $\Gamma$ are the respective irreducible vertices. The terms coupling the irreducible three-particle vertex to the full two-particle vertex account for diagrams reducible in a two-particle channel excluded from $\Gamma$, but three-particle irreducible. $F'$ denotes the three-particle vertex $1PI$ with respect to the one-particle channel associated with the three-particle channel in which we are solving the Bethe-Salpeter like equation. For the $pph$-like equation above, $F'$ is given by:
\begin{equation}
F' \begin{pmatrix} a & b & c \\ d & e & f \end{pmatrix} = F \begin{pmatrix} a & b & c \\ d & e & f \end{pmatrix} - \sum_1 F \begin{pmatrix} a & b \\ f & 1 \end{pmatrix} G(1) F \begin{pmatrix} 1 & c \\ d & e \end{pmatrix} . 
\end{equation}
\IPR \ contributions in the channel associated with the \IIIPR \ channel need to be removed from the Bethe-Salpeter-like equation because we prefer to work with dressed one-particle propagators which are difficult to categorise with respect to three-particle reducibility. Some self-energy diagrams can be cut into two parts by cutting three internal propagator lines\footnote{For quartic interaction, all $\Sigma$-diagrams are \IIIPR \ while for models with higher order interactions only some are.}. A proper treatment of these contributions would call for additional classification of the self-energy with respect to three-particle reducibility. It seems advantageous to circumvent these issues by discarding potentially problematic diagrams in the first place.

\subsection{\IIIPR \ in more than one channel}
\label{multi3PRsec}
\begin{figure}[H]
\centering
\includegraphics[width=0.85\linewidth]{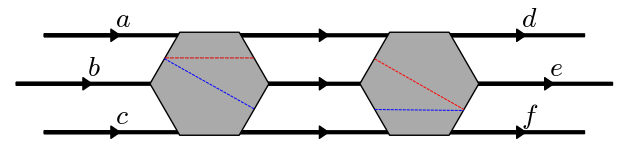}
\caption{\label{multi3PR} Possibilities of further decomposing a diagram with $a,b,c$ already separated from $d,e,f$. The red and blue lines give the possibilities of further decompositions.}
\end{figure}

A given diagram can be \IIIPR \ in any combination of two channels, there are no incompatible channels. Any diagram which is \IIIPR \ in more than one channel puts some restrictions on its decomposition. Let us investigate the possibilities of multiple reducibilities starting from a case as it is depicted in figure \ref{3P3PR}. We will discuss a diagram which is \IIIPR \ in more than one way. Without loss of generality, we can assume one of the channels to be the channel disconnecting $a,b,c$ from $d,e,f$, as depicted (This channel is unique in being the only $ppp$-like channel, but the further discussion does not depend on that fact). For the second channel, we know for sure that it will further disconnect the groups of three frequencies each into a pair of frequencies and a single frequency. For the following discussion, we will simply assume that the other channel is the one disconnecting $b,c,f$ from $a,d,e$. Let us now discuss the possibilities of disconnecting $a$ from $b,c$ as well as $f$ from $d,e$ using three or less cuts starting from a configuration as it is given in figure \ref{3P3PR}. The two possibilities on how the diagram needs to be further decomposed are given in figure \ref{multi3PR}.
We can  {use the red, dashed cuts in figure \ref{multi3PR}. The first (left) red cut disconnects  $a$ and one of the unlabeled lines in the middle of the figure (for aesthetic reasons we choose the uppermost one, but they are equivalent) from $b$,$c$ and the other two unlabeled lines. 
This requires cutting two Green's function lines in the left three-particle vertex (separating out a two-particle vertex). The second (right) red cut disconnects the other two unlabeled lines and $f$ from $d$, $e$, and the first unlabeled line. 
This second disconnection must then be made by a single cut (to have three cuts in total).
Altogether we have separated $a$, $d$, $e$ from $b$, $c$, $f$.
The other possibility of multiple \IIIPR \ is obtained by exchanging the cut(s) in the left and right three-particle vertex in figure \ref{multi3PR} and is depicted with blue, dashed lines. }

There is a 
further possibility of non-simultaneous reducibility, if both left and right vertex are one particle reducible. Then, by performing both tilted cuts as well as cutting the central line from the original decomposition, the diagram can be disconnected. We can gain valuable insights from the discussion. The constituent vertices of the first decomposition need to be \IPR \ and \IIPR \ respectively for a diagram to be \IIIPR \ in more than one channel. This implies that by getting rid of all \IPR \ and \IIPR \ contributions to the three-particle vertex prior to treating \IIIPR, we also preclude instances of multiple \IIIPR \ from occuring, reducing the combinatorical complexity in calculating the fully irreducible three-particle vertex.
\section{Examplary low-order, fully irreducible contributions to the three-particle vertex}
In this section, the lowest order (in the interaction) fully irreducible contributions to the three-particle vertex for a system with quartic interaction are derived. The lowest order at which such terms can appear turns out to be $6$. In the following, we will systematically derive the possible structures of such diagrams. To reduce the combinatorical complexity, we adopt notation symmetrised with respect to outgoing and incoming lines. To avoid confusion, the diagrammatic element corresponding to the interaction will be called interaction point. The usual name for the diagramatic representation of an interaction would be (bare interaction) vertex, which is an unfavourable choice in this context. Also, we will diregard directionality of the propagators. Due to the quartic interaction, each interaction point couples to exactly four propagator lines. Each propagator line couples to either two interaction points (if it is an internal line), or one interaction point (if it is an external line).
\\We first establish a lower boundary for the order of (one- and) two-particle irreducible, connected three-particle diagrams. Assume that the order of such a diagram was lower than $6$. Since there are $6$ external lines, at least one interaction point couples to at least $2$ external lines. If a point couples to $3$ external lines, the diagram is \IPR --by cutting the fourth line connected to it, the diagram disconnects. We must discard  such \IPR  \  contributions. If a point couples to two external lines, cutting the remaining two lines connected to it disconnects the diagram into two parts, it is therefore \IIPR . We conclude that a connected three-particle diagram that is one- and two-particle irreducible has to be at least of order $6$; each external line has to be connected to its own interaction point.

\begin{figure}
\centering
\begin{minipage}[b]{0.45\linewidth}
\includegraphics[width=0.9\textwidth]{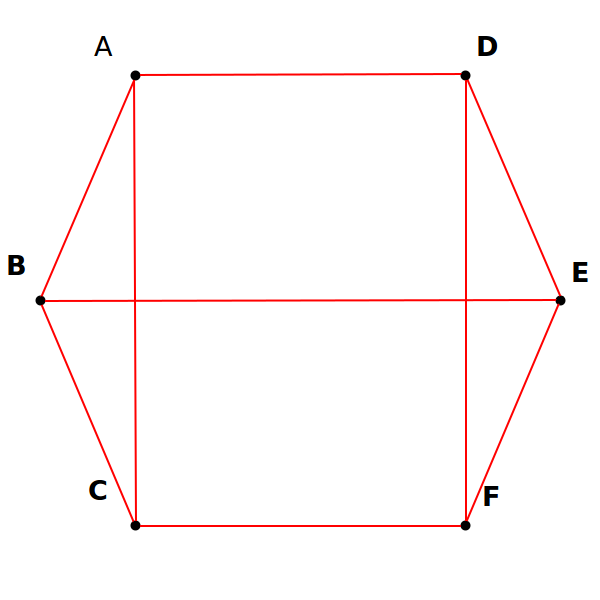}
\caption{\label{low3Triang}A low order three-particle diagram containing a triangle becomes three-particle reducible.}
\end{minipage}
\quad
\begin{minipage}[b]{0.45\linewidth}
\includegraphics[width=0.9\textwidth]{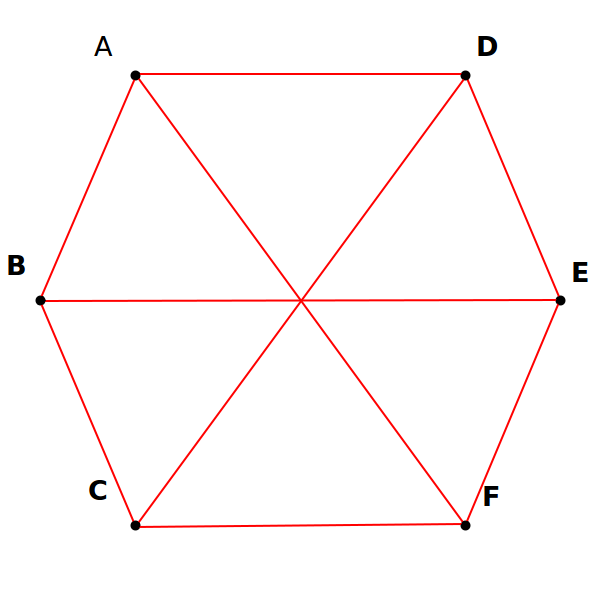}
\caption{\label{low3Hexa}The lowest-order contribution to the fully irreducible three-particle vertex, considering quartic interaction.}
\end{minipage}
\end{figure}

We now show that two interaction points that both connect to external lines  cannot be connected by more than one line. 
Assume that a pair of points is connected via a pair of lines. This pair of points is connected to two external lines as well as to the remaining diagram by another two lines--cutting those disconnects the diagram and therefore it is \IIPR .
\\We name the interaction points associated with (connected to) the external lines A,B,C,D,E and F. Without loss of generality, we can assume that B is connected to A, C and E. (Point B has to be connected to exactly three other points and we can choose the names freely.)
\\Next we exclude any triangles-triplets of points which are mutually connected to each other. Assuming a diagram which contains a triangle, we can name the points forming it A, B and C. This is achieved by connecting the points A and C, see figure \ref{low3Triang}. Once this is done, the remaining connections have to be performed as in figure \ref{low3Triang}\footnote{The choice of connecting A to D and C to F or A to F and C to D remains, but those diagrams are topologically equivalent.} Visual inspection of figure \ref{low3Triang} immediately uncovers the three-particle reducibility of the diagram. Hence, we discard these diagrams as well.
\\If formation of triangles is not admissible, because it would cause the resulting diagrams to become \IIIPR, point A cannot be connected to C or E and therefore has to be coupled with D and F. Point C has to be connected D and F as well. In total, we end up with a diagram like figure \ref{low3Hexa}, being the only type of sixth-order contribution to the fully irreducible three-particle vertex.

\section{Algorithm for prospective calculation of the fully irreducible three particle vertex}
In this section, an algorithm for calculating the fully (one-, two-, and three-particle) irreducible three-particle vertex is outlined. {We assume as a starting point the knowledge of the full three-particle vertex $F$ that can be calculated e.g.\ from the three-particle Green's function after removing disconnected diagrams.  We will then proceed as follows: first, all \IPR \ contributions will be removed from the three-particle vertex}, followed by any \IIPR \ ones. With all remaining diagrams being one- and two-particle irreducible, it is possible to solve three-particle Bethe-Salpeter-like equation for all diagrams reducible exclusively in a single three-particle channel, with the new three-particle reducible channels being completely disjunct, i.e. not containing any common diagrams. Knowledge about the decomposition of the two-particle vertex will be required. Let us now outline this algorithm in detail:

Going further into detail,
 the \IPR \ contributions can be easily removed as their structure is known. No three-particle diagram can be \IPR \ in more than one channel, but some \IPR \ diagrams are \IIPR \ as well. We can systematically remove all \IPR \ contributions by subtracting suitable combinations of full two-particle vertices connected by a single Green's function line. 
The \IPR \ vertex $\Phi_{1,abd-cef}$ in the channel $abd-cef$ is given by
\begin{equation}
\Phi_{1,abd-cef} \begin{pmatrix} a & b & c \\ d & e & f \end{pmatrix} = \sum_1 F \begin{pmatrix} a & b \\ d & 1 \end{pmatrix} \; G(1) \; F \begin{pmatrix} 1 & c \\ e & f \end{pmatrix}.
\end{equation}
The \IPR \ vertices in all $9$ one-particle channels can be subtracted from the full three-particle vertex $F$, yielding the fully one-particle irreducible (1PI) three particle vertex $\Gamma_{1\mathrm{PI}}$.
\begin{equation}
\Gamma_{1\mathrm{PI}} = F - \sum_{ijk-lmn} \Phi_{1,ijk-lmn} \; .
\label{eq:gamma_1PR}
\end{equation}
The above summation is performed over all $9$ \IPR \ channels $ijk-lmn$.

Since any \IPR \ diagrams were already removed from the full vertex, we have to take care not to subtract the ones which are \IIPR \ as well again. This means that we will establish Bethe-Salpeter-like equations for all one-particle irreducible diagrams $\Gamma_{1\mathrm{PI}}$. When doing so, not only do we need to remove all \IPR \ diagrams from the underlying set of diagrams, but also all the two-particle-reducible ones based on expanding \IPR \ diagrams. We will discuss the procedure of removing such diagrams for the $ab-de-cf$ {(pp -- pp -- ph)} and $ad-be-cf$ {(ph -- ph -- ph)} sets of compatible channels. For a given set of compatible two-particle channels, every one-particle reducible channel either disconnects the entering and leaving lines into two sets of \qt unpaired\qt \  variables, where no variables belonging to the same two-particle channel are on the same end of the diagram (class 1  \IPR \ diagrams), or into two sets of an \qt unpaired\qt \ variable and two \qt paired\qt \ ones (class 2  \IPR \ diagrams). For the $ab-de-cf$ set of channels, the completely \qt unpaired\qt \ (class 1)  \IPR \ channels are given by $adc-bef$, $adf-bce$, $ace-bdf$ and $aef-bcd$.

Any diagram \IPR \ in a completely \qt unpaired\qt \ channel is irreducible in all three two-particle channels. For the remaining channels, the \IPR , yet two-particle irreducible (in all three two-particle channels) contribution is given by the corresponding two-particle irreducible vertices, connected by a single one-particle Green's function. These \IPR \ contributions will be missing from our irreducible vertices, so any reducible diagrams built upon this irreducible basis need to be removed from the set of all diagrams for the Bethe-Salpeter-like equation as well.

\begin{figure}
\centering
\includegraphics[width=0.7\textwidth]{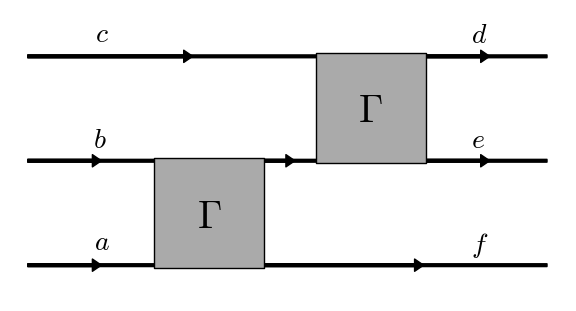}
\caption{\label{1PRfor2PR} \IPR \ (class 2  in the channel $abf-cde$ ) contribution to the two-particle $ab-de-cf$ irreducible vertex. The $\Gamma$'s are $pp$-irreducible two-particle vertices.}
\end{figure}

First we discuss the contributions from reducible diagrams built upon the \IPR \ diagrams made up from irreducible two-particle vertices, as depicted in figure \ref{1PRfor2PR}. To this basis diagram, full two-particle vertices are connected to $ab$, $de$ and $cf$ to recover all reducible diagrams in the corresponding channels. Connecting $F$ to $ab$ or $de$ yields all reducible two-particle vertices. Obviously, the resulting diagrams are all still one-particle reducible and were therefore already removed in the first step. Then further connecting another $F$ to $cf$ yields a diagram of the {triangle} structure in figure \ref{coollogo} above, with all three vertices being full two-particle vertices and the outer variables being grouped as $ab$, $de$ and $cf$. For this reason we need to remove these diagrams from the basis set of diagrams (the left hand side of our Bethe-Salpeter like equation). Note that this triangular diagram can be generated as a reducible contribution based on any class $2$ \IPR \ diagram  (of which there are either $5$ or $6$, for the $ph -ph-ph$ and $pp-pp-ph$ cases respectively). The issue arises because the decomposition of such diagrams into two-particle irreducible and reducible parts is not unique, i.e. there is more than one diagram one can end up with, depending on the order in which reducible contributions in the different channels are removed. Had we not manually taken care of these diagrams, an overcounting issue would have been the consequence.

\begin{figure}
\centering
\includegraphics[width=0.9\textwidth]{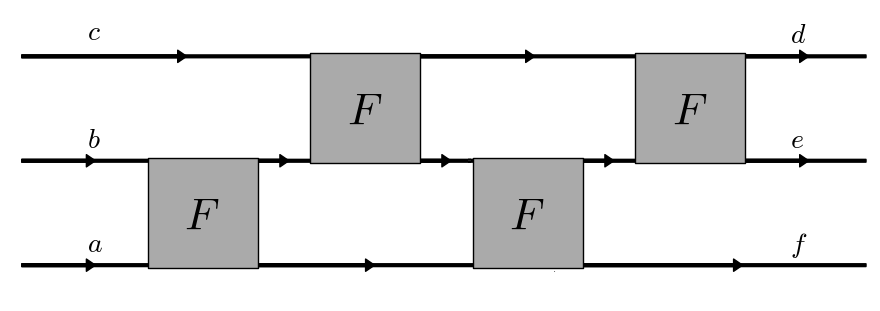}
\caption{\label{Butterfly} \IIPR \ diagrams in the channels $ab$ and $de$ built upon the \IPR \ diagrams in either the $acd-bef$, $ace-def$, $adf-bce$ or $adc-bef$ channel.}
\end{figure}

Now we proceed by investigating the behaviour of unpaired \IPR \ diagrams. To these diagrams, we connect any connection of either one, two or three two-particle vertices. Attaching a single two-particle vertex yields a triangle-diagram again, this time with one pairing of variables as in the two-particle channel. Therefore, also all triangles with one pair of variables fitting our two-particle channels needs to be removed from the l.h.s. of the {prospective Bethe-Salpeter-like} equation. Connecting two full two-particle vertices to the class $1$ \IPR \ term yields diagrams of the structure  depicted in figure \ref{Butterfly}. The \qt external\qt \ vertices have pairs of associated variables and there are three possible configurations of variables to be distributed between the external legs. Finally, the option of attaching all three vertices remains, generating diagrams which are reducible in all three two-particle channels. 

Triangle diagrams can be uniquely labeled by a set of compatible two-particle channels they are associated with. For $ph-ph-ph$ like sets of decompositions, two possible orientations for the inner line exist, which need to be taken into account. The $ad-be-cf$-triangle diagrams $T_{ad-be-cf}$ are given by
\begin{multline}
T_{ad-be-cf} \begin{pmatrix} a & b & c \\ d & e & f \end{pmatrix} = \sum_{1,2,3} G(1) G(2) G(3) \times 
\\ \left( - F \begin{pmatrix} a & 1 \\ d & 2 \end{pmatrix} F \begin{pmatrix} b & 2 \\ e & 3 \end{pmatrix} F \begin{pmatrix} c & 3 \\ f & 1 \end{pmatrix} - F \begin{pmatrix} a & 1 \\ d & 2 \end{pmatrix} F \begin{pmatrix} b & 3 \\ e & 1 \end{pmatrix} F \begin{pmatrix} c & 2 \\ f & 3 \end{pmatrix} \right) .
\end{multline}
The negative sign stems from a closed fermionic loop running around the inner part of the triangle diagram. For $pp-pp-ph$ cases there is only one possible orientation for the inner propagators and the triangle diagrams are given by
\begin{equation}
T_{ad-bc-ef} \begin{pmatrix} a & b & c \\ d & e & f \end{pmatrix} = \sum_{1,2,3} G(1) G(2) G(3) F \begin{pmatrix} a & 1 \\ d & 2 \end{pmatrix} F \begin{pmatrix} b & c \\ 1 & 3 \end{pmatrix} F \begin{pmatrix} 2 & 3 \\ e & f \end{pmatrix} .
\end{equation}
The diagrams with the structure as depicted in figure \ref{Butterfly} (which we will call "butterfly" diagrams) can be labeled by the pair of channels they are reducible in. They can be constructed based on triangle diagrams. The $ab-de$ butterfly diagrams $B_{ab-de}$ are given by
\begin{equation}
B_{ab-de} \begin{pmatrix} a & b & c \\ d & e & f \end{pmatrix} = \dfrac{1}{2} \sum_{1,2} G(1) G(2) F \begin{pmatrix} a & b \\ 1 & 2 \end{pmatrix}
 \left( T_{ac-bf-de} \begin{pmatrix} 1 & 2 & c \\ d & e & f \end{pmatrix} + T_{af-bc-de} \begin{pmatrix} 1 & 2 & c \\ d & e & f \end{pmatrix} \right) ,
\end{equation}
and the $ad-be$ ones by
\begin{equation}
B_{ad-be} \begin{pmatrix} a & b & c \\ d & e & f \end{pmatrix} = \dfrac{1}{1} \sum_{1,2} G(1) G(2) F \begin{pmatrix} a & 2 \\ 1 & d \end{pmatrix}
 \left( T_{ac-be-df} \begin{pmatrix} 1 & b & c \\ 2 & e & f \end{pmatrix} + T_{af-be-cd} \begin{pmatrix} 1 & b & c \\ 2 & e & f \end{pmatrix} \right) .
\end{equation}
It is in turn possible to express all of the diagrams generated from attaching three full vertices to a \IPR \ diagram, $H$, in terms of the butterfly diagrams
\begin{equation}
H_{ab-cf-de} \begin{pmatrix} a & b & c \\ d & e & f \end{pmatrix} = \dfrac{1}{1} \sum_{1,2} G(1) G(2) F \begin{pmatrix} c & 1 \\ 2 & f \end{pmatrix} B_{ab-de} \begin{pmatrix} a & b & 2 \\ d & e & 1 \end{pmatrix} .
\end{equation}
 To determine all \IIPR \ contributions to the one-particle irreducible three-particle vertex, we need to resolve the Bethe-Salpeter like equations
\begin{multline}
\Gamma_{1\mathrm{PI}}\begin{pmatrix} a & b & c \\ d & e & f \end{pmatrix} - \sum_{T, B, H} = \Gamma' \begin{pmatrix} a & b & c \\ d & e & f \end{pmatrix} 
 + \dfrac{1}{2} \sum_{1,2} \Gamma' \begin{pmatrix} a & b & c \\ 1 & 2 & f \end{pmatrix} G(1) G(2) \; F \begin{pmatrix} 1 & 2 \\ d & e \end{pmatrix} \\
 + \dfrac{1}{2} \sum_{1,2} \Gamma' \begin{pmatrix} 1 & 2 & c \\ d & e & f \end{pmatrix} G(1) G(2) \; F \begin{pmatrix} a & b \\ 1 & 2 \end{pmatrix} 
 + \dfrac{1}{1} \sum_{1,2} \Gamma' \begin{pmatrix} a & b & 1 \\ d & e & 2 \end{pmatrix} G(1) G(2) \; F \begin{pmatrix} c & 2 \\ 1 & f \end{pmatrix} \\
 + \dfrac{1}{4} \sum_{1,2,3,4} \Gamma' \begin{pmatrix} 1 & 2 & c \\ 3 & 4 & f \end{pmatrix} G(1) G(2) G(3) G(4) \; F \begin{pmatrix} a & b \\ 1 & 2 \end{pmatrix} F \begin{pmatrix} 3 & 4 \\ d & e \end{pmatrix}\\
 + \dfrac{1}{2} \sum_{1,2,3,4} \Gamma' \begin{pmatrix} 1 & 2 & 3 \\ d & e & 4 \end{pmatrix} G(1) G(2) G(3) G(4) \; F \begin{pmatrix} a & b \\ 1 & 2 \end{pmatrix} F \begin{pmatrix} c & 4 \\ 3 & f \end{pmatrix}\\
 + \dfrac{1}{2} \sum_{1,2,3,4} \Gamma' \begin{pmatrix} a & b & 1 \\ 3 & 4 & 2 \end{pmatrix} G(1) G(2) G(3) G(4) \; F \begin{pmatrix} c & 2 \\ 1 & f \end{pmatrix} F \begin{pmatrix} 3 & 4 \\ d & e \end{pmatrix}\\
 + \dfrac{1}{4} \sum_{1,2,3,4,5,6} \Gamma' \begin{pmatrix} 1 & 2 & 3 \\ 4 & 5 & 6 \end{pmatrix} G(1) G(2) G(3) G(4) G(5) G(6) \; F \begin{pmatrix} a & b \\ 1 & 2 \end{pmatrix} F \begin{pmatrix} 4 & 5 \\ d & e \end{pmatrix} F \begin{pmatrix} c & 6 \\ 3 & f \end{pmatrix}
\label{BS2PRpp}
\end{multline}
for $pp-pp-ph$ like separation and
\begin{multline}
\Gamma_{1\mathrm{PI}} \begin{pmatrix} a & b & c \\ d & e & f \end{pmatrix} - \sum_{T, B, H} = \Gamma' \begin{pmatrix} a & b & c \\ d & e & f \end{pmatrix} 
 + \dfrac{1}{1} \sum_{1,2} \Gamma' \begin{pmatrix} 1 & b & c \\ 2 & e & f \end{pmatrix} G(1) G(2) \; F \begin{pmatrix} a & 2 \\ 1 & d \end{pmatrix} \\
 + \dfrac{1}{1} \sum_{1,2} \Gamma' \begin{pmatrix} a & 1 & c \\ d & 2 & f \end{pmatrix} G(1) G(2) \; F \begin{pmatrix} b & 2 \\ 1 & e \end{pmatrix} 
 + \dfrac{1}{1} \sum_{1,2} \Gamma' \begin{pmatrix} a & b & 1 \\ d & e & 2 \end{pmatrix} G(1) G(2) \; F \begin{pmatrix} c & 2 \\ 1 & f \end{pmatrix} \\
 + \dfrac{1}{1} \sum_{1,2,3,4} \Gamma' \begin{pmatrix} 1 & 3 & c \\ 2 & 4 & f \end{pmatrix} G(1) G(2) G(3) G(4) \; F \begin{pmatrix} a & 2 \\ 1 & d \end{pmatrix} F \begin{pmatrix} b & 4 \\ 3 & e \end{pmatrix}\\
 + \dfrac{1}{1} \sum_{1,2,3,4} \Gamma' \begin{pmatrix} a & 1 & 3 \\ d & 2 & 4 \end{pmatrix} G(1) G(2) G(3) G(4) \; F \begin{pmatrix} b & 2 \\ 1 & e \end{pmatrix} F \begin{pmatrix} c & 4 \\ 3 & f \end{pmatrix}\\
 + \dfrac{1}{1} \sum_{1,2,3,4} \Gamma' \begin{pmatrix} a & 1 & 3 \\ b & 2 & 4 \end{pmatrix} G(1) G(2) G(3) G(4) \; F \begin{pmatrix} c & 2 \\ 1 & f \end{pmatrix} F \begin{pmatrix} 3 & 4 \\ d & e \end{pmatrix}\\
 + \dfrac{1}{1} \sum_{1,2,3,4,5,6} \Gamma' \begin{pmatrix} 1 & 3 & 5 \\ 2 & 4 & 6 \end{pmatrix} G(1) G(2) G(3) G(4) G(5) G(6) \; F \begin{pmatrix} a & 2 \\ 1 & d \end{pmatrix} F \begin{pmatrix} b & 4 \\ 3 & e \end{pmatrix} F \begin{pmatrix} c & 6 \\ 5 & f \end{pmatrix}
\label{BS2PRph}
\end{multline}
for $ph-ph-ph$ like separation. They provide a complete decomposition of the one-particle irreducible three-particle vertex in terms of reducibility in compatible two-particle channels. The $\Gamma'$ denote the vertices irreducible in all of the compatible channels associated with the equation ($ab$, $de$ and $cf$ for the first case and $ad$, $be$ and $cf$ for the second) and are different quantities in the equations above. The sums over $T, B, H$ are to include all triangle diagrams which share at least one two-particle channel with the Bethe-Salpeter like equation, all butterfly diagrams reducible in two channels appearing in the equation and the $H$ diagrams with all three channels being compatible.
There are $9$ equations equivalent to the first one and $6$ equivalent to the second. For each triplet of compatible chanels, the last line in the equations above gives the sum of all diagrams \IIPR \ in all three of them. Those sets of diagrams are completely disjunct, i.e. no diagram can be reducible in all three channels of more than one triplet. We define the vertices reducible exclusively in the triplets:
\begin{multline}
\Phi'_{ab,de,cf} \begin{pmatrix} a & b & c \\ d & e & f \end{pmatrix} = 
\\ \dfrac{1}{4} \sum_{1,2,3,4,5,6} \Gamma' \begin{pmatrix} 1 & 2 & 3 \\ 4 & 5 & 6 \end{pmatrix} G(1) G(2) G(3) G(4) G(5) G(6) \; F \begin{pmatrix} a & b \\ 1 & 2 \end{pmatrix} F \begin{pmatrix} 4 & 5 \\ d & e \end{pmatrix} F \begin{pmatrix} c & 6 \\ 3 & f \end{pmatrix} \\
 + T_{ab,de,cf} \begin{pmatrix} a & b & c \\ d & e & f \end{pmatrix} + H_{ab,de,cf} \begin{pmatrix} a & b & c \\ d & e & f \end{pmatrix}.
\end{multline}
\begin{multline}
\Phi'_{ad,be,cf}  \begin{pmatrix} a & b & c \\ d & e & f \end{pmatrix} = \\
\dfrac{1}{1} \sum_{1,2,3,4,5,6} \Gamma' \begin{pmatrix} 1 & 3 & 5 \\ 2 & 4 & 6 \end{pmatrix} G(1) G(2) G(3) G(4) G(5) G(6) \; F \begin{pmatrix} a & 2 \\ 1 & d \end{pmatrix} F \begin{pmatrix} b & 4 \\ 3 & e \end{pmatrix} F \begin{pmatrix} c & 6 \\ 5 & f \end{pmatrix} \\ 
+ T_{ad,be,cf}  \begin{pmatrix} a & b & c \\ d & e & f \end{pmatrix} + H_{ad,be,cf}  \begin{pmatrix} a & b & c \\ d & e & f \end{pmatrix}.
\end{multline}
Those quantities can be safely subtracted from the full vertex without incurring any overcounting issues. Also, we define the vertices \IIPR \ exclusively in pairs of compatible two-particle channels:
\begin{multline}
\Phi'_{ab,de} \begin{pmatrix} a & b & c \\ d & e & f \end{pmatrix} = 
\\ \dfrac{1}{4} \sum_{1,2,3,4} \Gamma' \begin{pmatrix} 1 & 2 & c \\ 4 & 5 & f \end{pmatrix} G(1) G(2) G(3) G(4) G(5) G(6) \; F \begin{pmatrix} a & b \\ 1 & 2 \end{pmatrix} F \begin{pmatrix} 4 & 5 \\ d & e \end{pmatrix} + B_{ab,de} \begin{pmatrix} a & b & c \\ d & e & f \end{pmatrix}.
\end{multline}
\begin{multline}
\Phi'_{ad,be}  \begin{pmatrix} a & b & c \\ d & e & f \end{pmatrix} = \\
\dfrac{1}{1} \sum_{1,2,3,4} \Gamma' \begin{pmatrix} 1 & 3 & c \\ 2 & 4 & f \end{pmatrix} G(1) G(2) G(3) G(4) G(5) G(6) \; F \begin{pmatrix} a & 2 \\ 1 & d \end{pmatrix} F \begin{pmatrix} b & 4 \\ 3 & e \end{pmatrix} + B_{ad,be}  \begin{pmatrix} a & b & c \\ d & e & f \end{pmatrix}.
\end{multline}
In this context exclusively \IIPR \ in a pair of channels includes all diagrams \IIPR \ in their respective pairs of channels, but irreducible in the third compatible channel. The exclusively reducible vertices in pairs of channels $\Phi'_{ij,kl}$ are related to the vertices \qt just\qt \ reducible in a pair of channels $\Phi_{ij,kl}$\footnote{Note that we restrict ourselves exclusively to one-particle irreducible vertices for the treatment of \IIPR .} via
\begin{equation}
\Phi_{ij,kl} = \Phi'_{ij,kl} + \Phi'_{ij,kl,mn},
\end{equation} 
with $ij$, $kl$, $mn$ being any triplet of compatible channels. The vertices exclusively \IIPR \ in pairs of channels do not share any diagrams with each other (if a diagram is reducible in three different two-particle channels it is instead included in $\Phi'_{ij,kl,mn}$ and no three-particle diagram can be \IIPR \ in more than $3$ channels.), nor with the vertices exclusively reducible in any triplet of compatible channels, so all of them can safely be subtracted from the full vertex.
\\Unfortunately, we cannot directly extract the vertex exclusively \IIPR \ in a single channel from any of the Bethe -Salpeter like equations above, because each \IIPR -channel appears in two triplets of compatible channels and information from both of the associated Bethe-Salpeter like equations is required to exclude all undesireable diagrams, i.e. those already included in some $\Phi'_{ij,kl}$ or $\Phi'_{ij,kl,mn}$. For example, the $\Gamma'$ calculated from equation \eqref{BS2PRpp} still contains contributions reducible in $df$ or $ce$.

We define new, simpler Bethe-Salpeter like equations
\begin{equation}
\Gamma_{1\mathrm{PI}}\begin{pmatrix} a & b & c \\ d & e & f \end{pmatrix} - \sum_T = \Gamma_{de} \begin{pmatrix} a & b & c \\ d & e & f \end{pmatrix} 
 + \dfrac{1}{2} \sum_{1,2} \Gamma_{de} \begin{pmatrix} a & b & c \\ 1 & 2 & f \end{pmatrix} G(1) G(2) \; F \begin{pmatrix} 1 & 2 \\ d & e \end{pmatrix}, 
\label{BS2PRppSingle}
\end{equation}
as well as
\begin{equation}
\Gamma_{1\mathrm{PI}}\begin{pmatrix} a & b & c \\ d & e & f \end{pmatrix} - \sum_T = \Gamma_{ad} \begin{pmatrix} a & b & c \\ d & e & f \end{pmatrix} 
 + \dfrac{1}{1} \sum_{1,2} \Gamma_{ad} \begin{pmatrix} 1 & b & c \\ 2 & e & f \end{pmatrix} G(1) G(2) \; F \begin{pmatrix} a & 2 \\ 1 & d \end{pmatrix} .
\label{BS2PRphSingle}
\end{equation}
For the $pp$ (\ref{BS2PRppSingle}) and $ph$-cases (\ref{BS2PRphSingle}) respectively, above equations provide decompositions into all diagrams reducible and irreducible in a single two-particle channel. The summations on the respective left hand sides are to be performed over all triangles including the channel in question. The reducible contributions
\begin{equation}
\Phi_{de} \begin{pmatrix} a & b & c \\ d & e & f \end{pmatrix} = \dfrac{1}{2} \sum_{1,2} \Gamma_{de} \begin{pmatrix} a & b & c \\ 1 & 2 & f \end{pmatrix} G(1) G(2) \; F \begin{pmatrix} 1 & 2 \\ d & e \end{pmatrix} + \sum_T
\end{equation}
and
\begin{equation}
\Phi_{ad} \begin{pmatrix} a & b & c \\ d & e & f \end{pmatrix} = \dfrac{1}{1} \sum_{1,2} \Gamma_{ad} \begin{pmatrix} 1 & b & c \\ 2 & e & f \end{pmatrix} G(1) G(2) \; F \begin{pmatrix} a & 2 \\ 1 & d \end{pmatrix} + \sum_T
\end{equation}
are not exclusively reducible in their respective channels. We have to remove any contributions which also appear in reducible contributions we have already calculated to arrive at exclusively reducible vertices $\Phi'$:
\begin{equation}
\Phi'_{de} = \Phi_{de} - \Phi'_{ab,de} - \Phi'_{de,cf} - \Phi'_{ab,de,cf} - \Phi'_{af,de} - \Phi'_{de,bc} - \Phi'_{af,de,bc} - \Phi'_{ac,de} - \Phi'_{de,bf} - \Phi'_{ac,de,bf}
\end{equation}
and
\begin{equation}
\Phi'_{ad} = \Phi_{ad} - \Phi'_{ad,be} - \Phi'_{ad,cf} - \Phi'_{ad,be,cf} - \Phi'_{ad,bf} - \Phi'_{ad,ce} - \Phi'_{ad,bf,ce} - \Phi'_{ad,bc} - \Phi'_{ad,ef} - \Phi'_{ad,bc,ef} .
\end{equation}
Once these quantities are recovered, all of the $\Phi'$ can simply be subtracted from the full three-particle vertex, removing all two-particle reducible contributions and yielding the fully one and two-particle irreducible three-particle vertex $\Gamma_{(1,2)\mathrm{PI}}$
\begin{equation}
\Gamma_{(1,2)\mathrm{PI}} = \Gamma_{1\mathrm{PI}} - \sum_{ij} \Phi'_{ij} - \sum_{\langle ij,kl \rangle} \Phi'_{ij,kl} - \sum_{\langle ij,kl,mn \rangle} \Phi'_{ij,kl,mn} , 
\end{equation}
where the summations are performed over all channels of two-particle reducibility, all distinct pairs of channels and all triplets of channels.

With all \IIPR \ diagrams gone, the \IIIPR \ ones remain to be removed. 

$\Gamma_{(1,2)\mathrm{PI}}$ contains all three-particle diagrams which are one- and two-particle irreducible, yet there are still \IIIPR \ contributions remaining. However, the task of removing such diagrams is made simpler by taking into account the discussion in section \ref{multi3PRsec}, reminding us that any diagram reducible in more than one three-particle channel is also reducible in at least one two-particle channel. Thus, by just determining the reducible contributions to $\Gamma_{(1,2)\mathrm{PI}}$ in all $10$ three-particle channels and subtracting them, we would recover the fully irreducible three particle vertex.

We have another look at the Bethe-Salpeter like equation \eqref{BS3pphMain} and compare two ways of writing it down:
\begin{multline}
F' \begin{pmatrix} a & b & c \\ d & e & f \end{pmatrix} = \Gamma^R \begin{pmatrix} a & b & c \\ d & e & f \end{pmatrix} 
 + \dfrac{1}{2} \sum_{1,2} \Gamma^R \begin{pmatrix} a & b & c \\ 1 & 2 & f \end{pmatrix} G(1) G(2) \; F \begin{pmatrix} 1 & 2 \\ d & e \end{pmatrix} \\
 + \dfrac{1}{1} \sum_{1,2} \Gamma^R \begin{pmatrix} a & b & 2 \\ d & 1 & f \end{pmatrix}  G(1) G(2) \; F \begin{pmatrix} 1 & c \\ e & 2 \end{pmatrix} 
 + \dfrac{1}{1} \sum_{1,2} \Gamma^R \begin{pmatrix} a & b & 2 \\ 1 & e & f \end{pmatrix}  G(1) G(2) \; F \begin{pmatrix} 1 & c \\ d & 2 \end{pmatrix} \\
 + \dfrac{1}{2} \sum_{1,2,3} \Gamma^R \begin{pmatrix} a & b & 3 \\ 1 & 2 & f \end{pmatrix}  G(1) G(2) G(3) \; F' \begin{pmatrix} 1 & 2 & c \\ d & e & 3 \end{pmatrix} 
\end{multline}
and 
\begin{multline}
F' \begin{pmatrix} a & b & c \\ d & e & f \end{pmatrix} = \Gamma^L \begin{pmatrix} a & b & c \\ d & e & f \end{pmatrix} 
 + \dfrac{1}{2} \sum_{1,2} F \begin{pmatrix} a & b \\ 1 & 2 \end{pmatrix} G(1) G(2) \; \Gamma^L \begin{pmatrix} 1 & 2 & c \\ d & e & f \end{pmatrix}   \\
 + \dfrac{1}{1} \sum_{1,2} F \begin{pmatrix} b & 2 \\ 1 & f \end{pmatrix}  G(1) G(2) \;   \Gamma^L \begin{pmatrix} a & 1 & c \\ d & e & 2 \end{pmatrix} 
 + \dfrac{1}{1} \sum_{1,2} F \begin{pmatrix} b & 2 \\ 1 & f \end{pmatrix}  G(1) G(2) \; \Gamma^L \begin{pmatrix} a & 1 & c \\ d & e & 2 \end{pmatrix}  \\
 + \dfrac{1}{2} \sum_{1,2,3} F' \begin{pmatrix} a & b & 3 \\ 1 & 2 & f \end{pmatrix}  G(1) G(2) G(3) \; \Gamma^L \begin{pmatrix} 1 & 2 & c \\ d & e & 3 \end{pmatrix} ,
\end{multline}
where we have defined 'left' and 'right' irreducible vertices $\Gamma^L$ and $\Gamma^R$. Both of them are irreducible in the three-particle (and one-particle) $abf-cde$ channel. While $\Gamma^R$ is additionally irreducible in $cd$, $ce$ and $de$, $\Gamma^L$ is in $ab$, $af$ and $bf$. We are interested in recovering all \IIIPR , yet one- and two-particle irreducible diagrams from above equations. All of the three-particle reducible vertex contributions are given by the terms where $F'$ couples to the respective $\Gamma$. We insert the expression for $F'$ extracted from one of the equations into the three-particle-reducible part of the other, recovering the reducible vertex $\Phi$:
\begin{multline}
\Phi \begin{pmatrix} a & b & c \\ d & e & f \end{pmatrix} = \dfrac{1}{4} \sum_{1,2,3,4,5,6} \Gamma^R \begin{pmatrix} a & b & 3 \\ 1 & 2 & f \end{pmatrix}  G(1) G(2) G(3) \; \cdot \; \\
 \Bigg( F' \begin{pmatrix} 1 & 2 & 6 \\ 4 & 5 & 3 \end{pmatrix} + F \begin{pmatrix} 1 & 2 \\ 4 & 5 \end{pmatrix} \delta_{3,6} G^{-1}(3) + 2 \; F \begin{pmatrix} 1 & 6 \\ 4 & 3 \end{pmatrix} \delta_{2,5} G^{-1}(2) + 2 \; F \begin{pmatrix} 2 & 6 \\ 5 & 3 \end{pmatrix} \delta_{1,4} G^{-1}(1) \\ + 2 \; \delta_{1,4} \, \delta_{2,5} \, \delta_{3,6}  \, G^{-1}(1) \, G^{-1}(2) \, G^{-1}(3) \Bigg) G(4) G(5) G(6) \; \Gamma^L \begin{pmatrix} 4 & 5 & c \\ d & e & 6 \end{pmatrix}.
\end{multline}
The quantity in the brackets can be calculated straightforwardly and will be abbreviated as effective vertex $V_{eff}$. From the reducible vertex, we want to remove all \IPR \ and \IIPR \ contributions. There are no \IPR \ contributions currently included. One can easily verify that any \IPR \ contributions to either $\Gamma^L$ or $\Gamma^R$ immediately lead to \IIPR \ contributions to $\Phi$. We therefore remove the remaining $8$ \IPR \ terms in $\Gamma^L$ and $\Gamma^R$, keeping in mind that they are two-particle irreducible in some channels. For $\Gamma^R$ the remaining \IPR \ terms, $\Gamma^R_{1\mathrm{PR}}$, are given by
\begin{multline}
\Gamma^R_{1\mathrm{PR}} = \sum_1 G(1) \cdot \Bigg( 
 F \begin{pmatrix}  a & b \\ d & 1 \end{pmatrix} \Gamma \begin{pmatrix} { \textcolor{green}{ 1}} & c \\ e & { \textcolor{green}{ f}}  \end{pmatrix} + 
F \begin{pmatrix}  a & b \\ 1 & e \end{pmatrix} \Gamma \begin{pmatrix} { \textcolor{green}{ 1}} & c \\ d & { \textcolor{green}{ f}}  \end{pmatrix} + \\
F \begin{pmatrix}  b & 1 \\ e & f \end{pmatrix} \Gamma \begin{pmatrix} { \textcolor{green}{ a}} & c \\ d & { \textcolor{green}{ 1}}  \end{pmatrix} + 
F \begin{pmatrix}  1 & b \\ f & d \end{pmatrix} \Gamma \begin{pmatrix} { \textcolor{green}{ a}} & c \\ { \textcolor{green}{ 1}} & e  \end{pmatrix} + 
F \begin{pmatrix}  a & c \\ 1 & f \end{pmatrix} \Gamma \begin{pmatrix} { \textcolor{green}{ 1}} & { \textcolor{green}{ b}} \\ d & e  \end{pmatrix} + \\
F \begin{pmatrix}  1 & a \\ e & f \end{pmatrix} \Gamma \begin{pmatrix} { \textcolor{green}{b}} & c \\ d & { \textcolor{green}{ 1}}  \end{pmatrix} + 
F \begin{pmatrix}  1 & a \\ f & d \end{pmatrix} \Gamma \begin{pmatrix} { \textcolor{green}{ b}} & c \\ e & { \textcolor{green}{ 1}} \end{pmatrix} + 
F \begin{pmatrix}  b & c \\ 1 & f \end{pmatrix} \Gamma \begin{pmatrix} { \textcolor{green}{ a}} & { \textcolor{green}{ 1}} \\ d & e  \end{pmatrix}
\Bigg).
\end{multline}
Colours were used to denote inseperable outer legs of the two-particle $\Gamma$,i.e.
$\Gamma \begin{pmatrix} { \textcolor{green}{ a}} & b \\ { \textcolor{green}{ c}} & d  \end{pmatrix}$ is the $ph$-irreducible vertex, $\Gamma \begin{pmatrix} { \textcolor{green}{ a}} & b \\ c & { \textcolor{green}{ d}}  \end{pmatrix}$ the $\overline{ph}$ one and the $pp$-irreducible vertex is given by $\Gamma \begin{pmatrix} { \textcolor{green}{ a}} & { \textcolor{green}{ b}} \\ c & d  \end{pmatrix}$.
We remove any \IPR \ terms from $\Gamma^L$ and $\Gamma^R$.
\begin{equation}
\Gamma'^{\,L} = \Gamma^L - \Gamma_{1\mathrm{PR}}^L
\end{equation}
\begin{equation}
\Gamma'^{\,R} = \Gamma^R - \Gamma^R_{1\mathrm{PR}}
\end{equation}
The last task remaining in determining a proper $\Gamma_{abf-cde}$ is either removing all $ab$, $af$ and $bf$-reducible terms from $\Gamma'^{\,R}$ or all $cd$, $ce$ and $de$ ones from $\Gamma'^{\,L}$ (which is equivalent), yielding $\Gamma_{eff}$.
\begin{multline}
\Gamma'^{\,R} \begin{pmatrix} a & b & c \\ d & e & f \end{pmatrix} = \Gamma_{eff} \begin{pmatrix} a & b & c \\ d & e & f \end{pmatrix} 
 + \dfrac{1}{2} \sum_{1,2} F \begin{pmatrix} a & b \\ 1 & 2 \end{pmatrix} G(1) G(2) \; \Gamma_{eff} \begin{pmatrix} 1 & 2 & c \\ d & e & f \end{pmatrix}   \\
 + \dfrac{1}{1} \sum_{1,2} F \begin{pmatrix} b & 2 \\ 1 & f \end{pmatrix}  G(1) G(2) \;   \Gamma_{eff} \begin{pmatrix} a & 1 & c \\ d & e & 2 \end{pmatrix} 
 + \dfrac{1}{1} \sum_{1,2} F \begin{pmatrix} b & 2 \\ 1 & f \end{pmatrix}  G(1) G(2) \; \Gamma_{eff} \begin{pmatrix} a & 1 & c \\ d & e & 2 \end{pmatrix}  
\end{multline}
\begin{multline}
\Gamma'^{\,L} \begin{pmatrix} a & b & c \\ d & e & f \end{pmatrix} = \Gamma_{eff} \begin{pmatrix} a & b & c \\ d & e & f \end{pmatrix} 
 + \dfrac{1}{2} \sum_{1,2} \Gamma_{eff} \begin{pmatrix} a & b & c \\ 1 & 2 & f \end{pmatrix} G(1) G(2) \; F \begin{pmatrix} 1 & 2 \\ d & e \end{pmatrix} \\
 + \dfrac{1}{1} \sum_{1,2} \Gamma_{eff} \begin{pmatrix} a & b & 2 \\ d & 1 & f \end{pmatrix}  G(1) G(2) \; F \begin{pmatrix} 1 & c \\ e & 2 \end{pmatrix} 
 + \dfrac{1}{1} \sum_{1,2} \Gamma_{eff} \begin{pmatrix} a & b & 2 \\ 1 & e & f \end{pmatrix}  G(1) G(2) \; F \begin{pmatrix} 1 & c \\ d & 2 \end{pmatrix} 
\end{multline}
With $\Gamma_{eff}$ available, we are able to express the exclusively \IIIPR \  vertex, $\Phi'_{abf-cde}$ as
\begin{multline}
\Phi'_{abf-cde} \begin{pmatrix} a & b & c \\ d & e & f \end{pmatrix} = \dfrac{1}{4} \sum_{1,2,3,4,5,6} \Gamma_{eff} \begin{pmatrix} a & b & 3 \\ 1 & 2 & f \end{pmatrix}  G(1) G(2) G(3) \\ 
V_{eff} \begin{pmatrix} 1 & 2 & 6 \\ 4 & 5 & 3 \end{pmatrix} G(4) G(5) G(6) \; \Gamma_{eff} \begin{pmatrix} 4 & 5 & c \\ d & e & 6 \end{pmatrix}.
\end{multline}
Applying the above procedure for all $10$ channels (actually only the $pph$ and $ppp$ case are independent), all remaining reducible contributions to $\Gamma_{(1,2)\mathrm{PI}}$ can be eliminated, yielding the fully irreducible three-particle vertex $\Lambda$
\begin{equation}
\Lambda = \Gamma_{(1,2)\mathrm{PI}} - \sum_{\langle ijk-lmn\rangle} \Phi'_{ijk-lmn}.
\end{equation}
The summation is again performed over all $10$ channels of three-particle reducibility.
\newpage

\section{Conclusion}

The issue of irreducibility on the three-particle level is much more involved than on the two-particle level.
Let us here sum up the different definitions of one-, two- and three-particle reducibility we used and
how they are mutually exclusive or non-exclusive:

\begin{itemize}
\item 
A connected three-particle diagram is \IPR \ if it can be cut into two disconnected two-particle parts, each with three of the original six outer legs by removing a single internal one-particle propagator. There are 9 channels of one-particle reducibility. The channels are labeled by the pair of triplets of outer legs which remain connected to each other when performing the cut. A diagram can be \IPR \ in one channel at most. 
\item
A connected three-particle diagram is \IIPR \ if it can be cut into two disconnected parts, a three-particle and a two-particle one, by removing two internal one-particle propagators. The three-particle part remains connected to four of the original six outer legs and the two-particle part to the remaining two outer legs. There are 15 channels of two-particle reducibility. We have labeled the channels by the pair of outer legs which is disconnected from the remaining diagram. Two-particle channels are compatible if a single diagram can be reducible in all of them. There are 30 pairs of compatible channels and 10 triplets of compatible channels {(Here,  a diagram is said to be simultaneously \IIPR \  in $n$ channels if it can be disconnected into $n$ two-particle parts with the corresponding pairs of the outer legs and a remaining three-particle part by removing $2n$ internal one-particle propagators.)}. A diagram can be \IPR \ and \IIPR \ in two-particle channels which disconnect a pair of outer legs that remains together in the \IPR \ decomposition.
\item
A connected three-particle diagrams is \IIIPR \ if it can be cut into two three-particle parts each of which remains connected to three external legs by removing three internal one-particle propagators. There are 10 different channels of three-particle reducibility. The channels are labeled by the pair of triplets of outer legs which remain connected to each other when disconnecting the diagram. A given diagram can be \IIIPR \ in more than one channel. If a diagram is \IIIPR \ in more than one channel, it is also \IIPR . A diagram can be \IPR \ and \IIIPR , but only in the channels which are labeled the same. A diagram can be both \IIPR \ and \IIIPR .
\end{itemize}

Using this insight and proper subtraction, we have derived the  equations for determining, in principle, the three-particle vertex that  is irreducible in a given channel  as well as the fully irreducible three-particle vertex. We have not derived parquet--like equations that use the latter as a starting point and obtain from it the full three-particle vertex (as well as two- and one-particle vertices). 
In the main text we have employed an approximation which is based on { simple \IIPR \ ladders} for being actually able to do some calculations on the three-particle level. We added this Supplemental Information since we think it might be 
a helpful framework for future  more involved three-particle vertex calculations.

